\documentclass[prl,aps,twocolumn,floatfix,preprintnumbers]{revtex4-2}
\usepackage[utf8]{inputenc}
\usepackage{graphicx,psfrag}
\usepackage{mathrsfs}
\usepackage{amsmath,amsfonts,amssymb}
\usepackage{multirow, rotating}
\usepackage{textgreek}
\usepackage{diagbox}
\usepackage{comment}
\usepackage{xcolor}
\usepackage{enumerate}
\usepackage{booktabs,bbm}
\usepackage{siunitx}

\usepackage{hyperref}
\usepackage{cleveref}
\hypersetup{
    colorlinks = true,
    linkcolor = {blue},
    citecolor = {blue},
    urlcolor = {blue},
    linkbordercolor = {white},
    }

\usepackage{color}
\definecolor{cyan}{rgb}{0,0.9,0.9}
\definecolor{orange}{rgb}{0.9,0.5,0}
\definecolor{magenta}{rgb}{1,0,1}
\definecolor{purple}{rgb}{0.8,0.4,0.8}
\definecolor{gray}{rgb}{0.8242,0.8242,0.8242}
\definecolor{mgreen}{rgb}{0.1,0.8,0.1}

\usepackage[normalem]{ulem}

\newcommand{\ve}{V_{E,\mathbbm{1}}}
\newcommand{\mve}{$\ve$}
\renewcommand{\bf}{\mathcal{B}^{V_{E,\mathbbm{1}}}_{\rm TPE}}

\newcommand{\dift}{{\rm d}}

\begin{document}
\preprint{LA-UR-22-21458}

\title{Revealing the strength of three-nucleon interactions with the Einstein Telescope}

\author{Henrik \surname{Rose}$^{1}$}
\author{Nina \surname{Kunert}$^{1}$}
\author{Tim \surname{Dietrich}$^{1,2}$}
\author{Peter~T.~H. \surname{Pang}$^{3,4}$}
\author{Rory~\surname{Smith}$^{5}$}
\author{Chris \surname{Van Den Broeck}$^{3,4}$}
\author{Stefano \surname{Gandolfi}$^{6}$}
\author{Ingo \surname{Tews}$^{6}$}

\affiliation{${}^1$Institute for Physics and Astronomy, University of Potsdam, D-14476 Potsdam, Germany}
\affiliation{${}^2$Max Planck Institute for Gravitational Physics (Albert Einstein Institute), Am M\"uhlenberg 1, D-14476 Potsdam, Germany}
\affiliation{${}^3$Nikhef, Science Park 105, 1098 XG Amsterdam, The Netherlands}
\affiliation{${}^4$Institute for Gravitational and Subatomic Physics (GRASP), Utrecht University, Princetonplein 1, 3584 CC Utrecht, The Netherlands}
\affiliation{${}^5$OzGrav: The ARC Centre of Excellence for Gravitational Wave Discovery, Clayton VIC 3800, Australia}
\affiliation{${}^6$Theoretical Division, Los Alamos National Laboratory, Los Alamos, NM 87545, USA}

\begin{abstract}

Three-nucleon forces are crucial for the accurate description of nuclear systems, including dense matter probed in neutron stars.
We explore nuclear Hamiltonians that reproduce two-nucleon scattering data and properties of light nuclei, but differ in the three-nucleon interactions among neutrons.
While no significantly improved constraints can be obtained from current astrophysical data, we show that observations of neutron star mergers by next-generation detectors like the proposed Einstein Telescope could provide strong evidence to distinguish between these Hamiltonians.

\end{abstract}

\flushbottom
\maketitle
{\it Introduction - }
Neutron stars (NSs) are among the most extreme objects in the universe~\cite{Lattimer:2000nx,Chamel:2008ca,Ozel:2016oaf,Gandolfi:2019zpj} and contain observable matter at the highest densities realized anywhere in nature. 
Inside NSs, densities up to several times the nuclear saturation density, corresponding to $\rho_{\rm sat}\approx \SI{2.7e14}{\gram\per\cubic\cm}$ can be reached.
However, the structural properties of typical NSs, i.e., their masses, radii, and deformabilities, are determined to a large extent by dense matter up to $2-3\,\rho_{\rm sat}$.
At these densities, neutron-star matter consists mainly of neutrons and protons whose microscopic interactions determine the macroscopic properties of NSs. 
The macroscopic NS properties can, in turn, be extracted from analyses of data from astrophysical observations, for example gravitational wave (GW)~\cite{LIGOScientific:2017qsa,Abbott:2018exr,Abbott:2018wiz} and electromagnetic (EM) signals~\cite{Bauswein:2017vtn,Radice:2017lry,Most:2018hfd,Coughlin:2018fis,Dietrich:2020efo} from NS mergers, or EM observations of isolated NSs, e.g., from the Neutron star Interior Composition Explorer (NICER)~\cite{Riley:2019yda,Miller:2019cac,Riley:2021pdl,Miller:2021qha}.
Hence, by comparing predictions of theoretical models for dense nuclear matter and astrophysical data on typical neutron stars, one can infer properties of microscopic nuclear interactions.

In the previous decade, tremendous progress has been made in calculating the properties of nuclear systems from microscopic nuclear theory.
This progress was driven mainly by the development of systematic interactions from chiral effective field theory (EFT)~\cite{Epelbaum:2008ga,Machleidt:2011zz} as well as improvements to many-body computational methods.
These methods solve the many-body Schr{\"o}dinger equation numerically for a system described by a nuclear Hamiltonian that describes the kinetic energy of the particles as well as their interactions, $\mathcal{H}=T+V_{NN}+V_{3N}+\cdots$, where $V_{NN}$ describes two-nucleon (NN) interactions, $V_{3N}$ describes three-nucleon (3N) interactions, and the dots indicate additional many-body forces.
Calculations of properties of atomic nuclei and isotopic chains~\cite{Otsuka:2009cs,Wienholtz:2013nya}, and studies of nuclear matter~\cite{Day:1983gga,Drischler:2017wtt,Lonardoni:2020} have shown that 3N interactions are an important ingredient in nuclear Hamiltonians and crucial to accurately describe data.
In chiral EFT, 3N interactions are usually constructed to reproduce properties of light nuclei~\cite{Navratil:2009ut,Hebeler:2013nza,Lynn:2017fxg} and then used to study heavier atomic nuclei and neutron-rich matter relevant for astrophysics. 
The latter requires the extrapolation of these interactions from nearly symmetric to almost pure neutron systems, which might suffer from systematics if interactions among neutrons are poorly constrained.
Hence, it is desirable to investigate if one can constrain these interactions directly in neutron-rich systems.
In this letter, we examine how well we can distinguish between nuclear Hamiltonians that include different 3N interactions by ana\-lyzing GW signals of NS mergers, fully taking into account present uncertainties in nuclear theory.
We probe how different tidal properties, due to the different 3N contributions, can be extracted from a catalogue of synthetic signals as observed in future third-generation detectors, e.g., the Einstein Telescope (ET)~\cite{Punturo:2010zz,Hild:2010id}.

{\it Equations of state for different three-nucleon interactions - }
To analyze the impact of 3N interactions on the equation of state (EOS) of NSs, we follow Ref.~\cite{Tews:2018iwm} and construct two EOS sets constrained by auxiliary field diffusion Monte Carlo calculations~\cite{Schmidt:1999lik,Carlson:2014vla,Lynn:2019rdt} of pure neutron matter for two local Hamiltonians from chiral EFT~\cite{Gezerlis:2013ipa,Gezerlis:2014zia,Lynn:2015jua}.
These Hamiltonians differ in their 3N interactions in pure neutron matter (the interactions of Ref.~\cite{Lynn:2015jua} named TPE and $V_{E,\mathbbm{1}}$~\footnote{We do not investigate the $V_{E,\tau}$ interaction of Ref.~\cite{Lynn:2015jua} because it leads to negative pressure in pure neutron matter below 2$\,\rho_{\rm sat}$.}), but they give a similar description in atomic nuclei~\cite{Lonardoni:2018nob}.

The difference in the neutron-matter description originates from regulator artifacts in the EOS due to the 3N contact interaction $V_{E}$~\cite{Lynn:2015jua}.
In pure neutron matter, without any regulators, the two-pion--exchange (TPE) interaction is the only 3N contribution because the shorter-range one-pion-exchange--contact interaction $V_D$ and 3N contact $V_E$ vanish due to their spin-isospin structure and the Pauli principle, respectively~\cite{Hebeler:2009iv}.
However, when local regulators are applied, the contact interactions acquire a finite range and start to contribute also to pure neutron systems~\cite{Lynn:2015jua,Huth:2017wzw}.
Here, we use these regulators artifacts to our advantage to test the sensitivity of the EOS to different 3N interactions~\footnote{In principle, these regulator artifacts can be thought of as sub-leading 3N contact interactions, appearing first at N$^4$LO in chiral EFT.}.
The first Hamiltonian only contains the TPE interaction, while the second Hamiltonian additionally contains a repulsive 3N contact piece with the identity operator, $V_{E,\mathbbm{1}}$. 
For both Hamiltonians, we calculate the EOS up to $2\,\rho_{\rm sat}$, estimate the truncation uncertainties according to the description used in Ref.~\cite{Lynn:2015jua}, and extend it to higher densities using the speed-of-sound extrapolation scheme introduced in Ref.~\cite{Tews:2018iwm}.
For the extension, the prior in the radius of a typical $1.4M_{\odot}$ NS is ``natural'', i.e., we directly use the generated EOS as prior and do not post-select EOS to generate a certain prior shape.
Hence, both sets enable us to explore the impact of different 3N interaction strengths while taking into account all theoretical uncertainties.
For simplicity, we refer to the two sets as TPE and \mve, too.

Including these uncertainties is key in answering the question of whether current and future observations can distinguish between nuclear Hamiltonian and, in our case, can reveal the strength of 3N interactions, an important difference to Refs.~\cite{Maselli:2020uol,Sabatucci:2022qyi}, who were the first to investigate the impact of the GW measurements with current and future GW detector generations on 3N forces in nuclear Hamiltonian. 
First, uncertainties in the nuclear EOS are not solely originating from unknown 3N interactions, which is reflected by the truncation uncertainty separately estimated for each Hamiltonian employed here.  
Second, at higher densities in the core of NSs, a description in terms of nucleonic degrees of freedom alone might fail as exotic forms of matter might appear.
Using the speed-of-sound extrapolation allows us to account for that.
Both uncertainties soften the constraining power of multimessenger data but are crucial to make robust statements about prospects of constraining nuclear Hamiltonians.

{\it Injection campaign - }
Due to the tidal deformation of the stars during the inspiral phase, the EOS leaves a characteristic imprint on the observable gravitational waveform observed in NS mergers~\cite{Hinderer:2007mb,Chatziioannou:2020pqz}. 
Using the EOS as a sampling parameter, we study the resulting effects in a nested sampling approach to parameter estimation~\cite{Skilling:2006gxv,Thrane:2018qnx}. 
This aims at computing the evidence $\mathcal{Z}$ which normalizes the posterior distribution of the parameter space in a Bayesian framework.
We follow common practice to express model preference for the TPE or \mve\ Hamiltonians by a Bayes factor $\bf=\frac{\mathcal{Z}_{\ve}}{\mathcal{Z}_{\rm TPE}}$
or its logarithm $\ln \bf= \ln\mathcal{Z}_{\ve} -\ln \mathcal{Z}_{\rm TPE} $.
The EOS sampling prior is weighted conservatively by incorporating a lower bound on the TOV mass in agreement with precise pulsar observations~\cite{Antoniadis:2013pzd,Arzoumanian:2017puf,Fonseca:2021wxt} (compare Fig. S1 of Ref.~\cite{Dietrich:2020efo}).
We treat each GW signal as an independent event connected by the EOS as the only hyperparameter.
Correspondingly, the Bayes factor after $N$ detections is the product of each individual event's Bayes factor: $\bf=\prod_{k=1}^N {\bf}^{\,k}$.

A reanalysis of the GW transient GW170817 with respect to 3N interactions proves uninformative (see supplemental material for details).
In principle, the EOS is linked to EM observables, too, as it determines NS radii and, thus, affects the properties of ejected matter.
Intricate models of this connection to EM counterparts are under development, but current uncertainties do not allow stringent constraints on 3N interactions.
These modelling efforts will profit from additional multimessenger events observed with the present detector generation~\cite{VIRGO:2014yos,LIGOScientific:2014jea}.
Yet event rates are highly uncertain and kilonova rates are especially poorly constrained~\cite{LIGOScientific:2021djp,LIGOScientific:2021usb,Petrov:2021bqm,Colombo:2022zzp}. 
Therefore, significantly improved constraints may not be expected before future detector technology becomes operational~\cite{Pacilio:2021}.
In the GW sector, this refers to the third-generation Einstein Telescope (ET)~\cite{Punturo:2010zz} in Europe and the proposed Cosmic Explorer (CE)~\cite{Reitze:2019iox} in the US.
We, therefore, analyze synthetic GW signals detected with ET. 
To that end, we choose an example EOS for both Hamiltonians and perform a volume-limited injection study based on 20 synthetic systems from a realistic binary NS population~\cite{Ozel:2016oaf}.
For each injected EOS, we compare Bayesian parameter estimation over the TPE and \mve\ set, amounting to a total of 80 inference runs in the frequency range \SI{30}{Hz} to \SI{2048}{Hz}.
 
\begin{figure*}[t]
    \centering
    \includegraphics[width=0.49\textwidth]{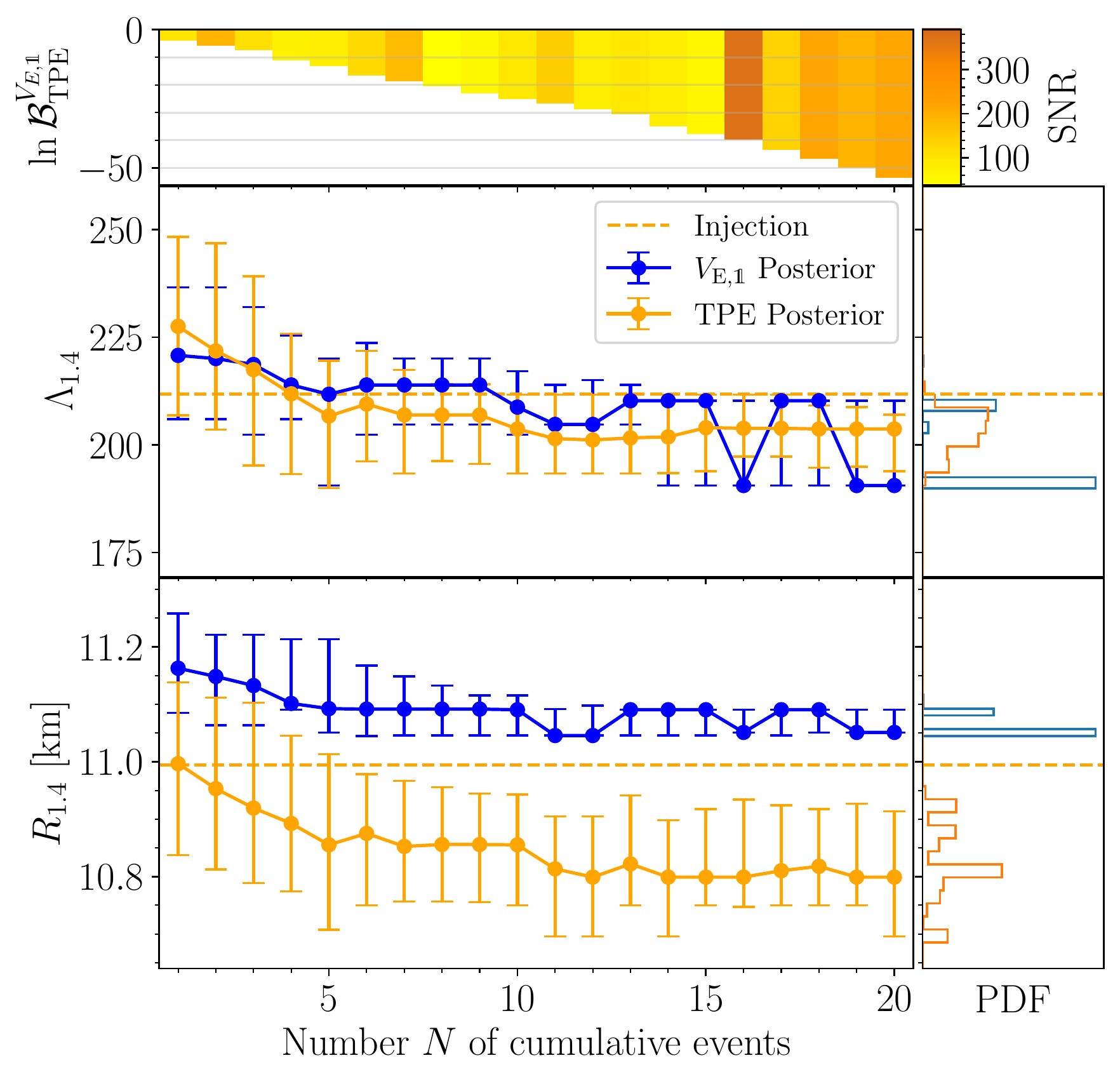}
    \includegraphics[width=0.49\textwidth]{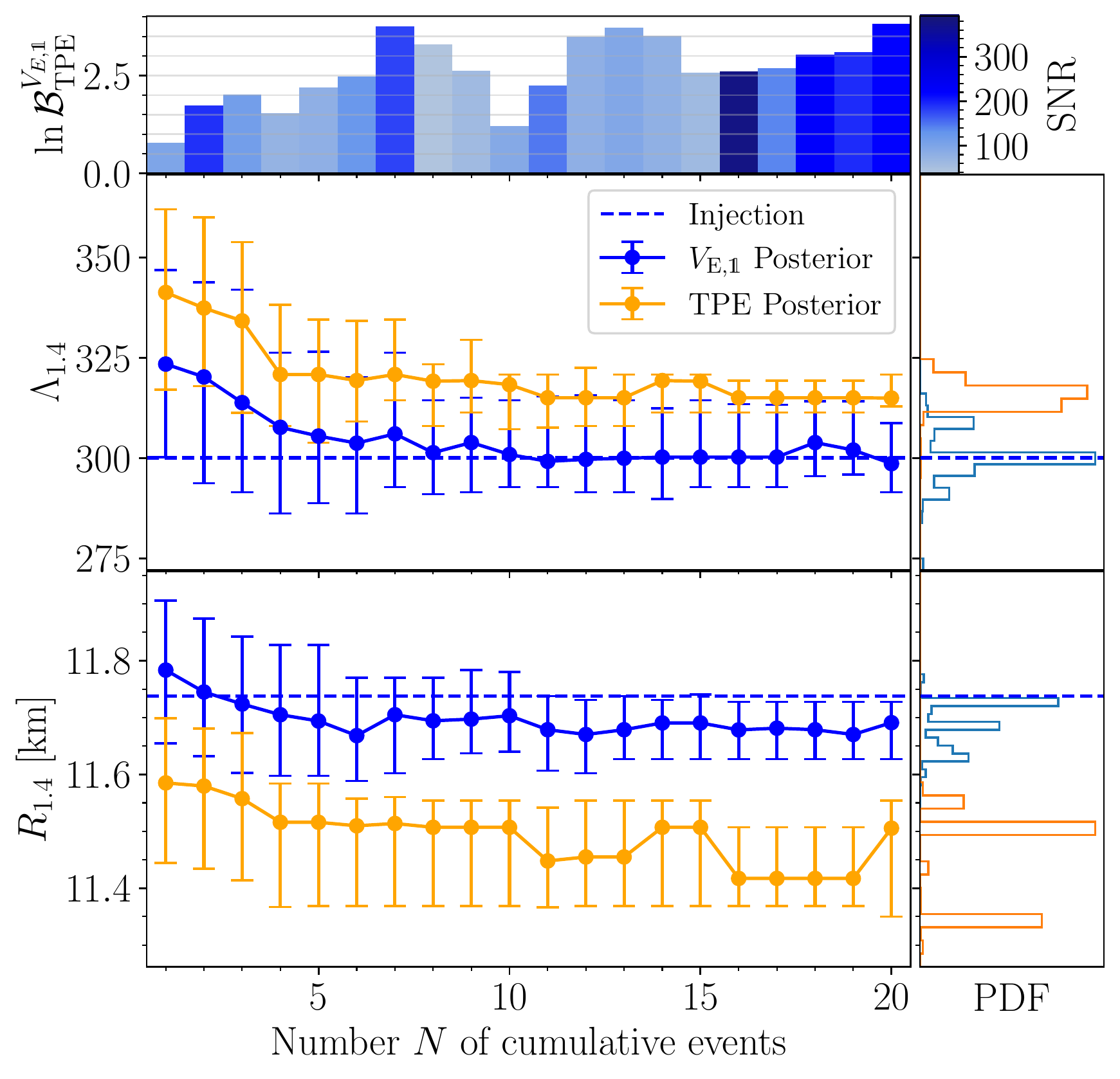}
    \caption{\textbf{Joint EOS posteriors and Bayes Factors:} 
    For a TPE (left) and \mve\ (right) injection, we show model preference as expressed by a cumulative Bayes factor (top), color-coded by the event's signal-to-noise ratio (SNR).
    While the evidence in favor of a TPE injection piles up almost linearly, some runs with low SNR disfavor the \mve\ set when injected.
    Joint posteriors on tidal deformability \textLambda\ (middle) and radius $R$ (bottom) of a \SI{1.4}{M_\odot} NS are displayed as a function of total signals observed in random order and as a probability density function (PDF) after all 20 events. }
    \label{fig:main_result}
\end{figure*}

{\it Results - }
Combining the additional information from each event into a cumulative EOS posterior, we obtain the results illustrated in Fig.~\ref{fig:main_result}.
On the left, we show the EOS posterior for both EOS sets when injecting a TPE EOS. 
Evidence in favor of the TPE Hamiltonian accumulates very quickly and essentially independent of further system parameters (top).
Particularly, we see no correlation with the color-coded signal-to-noise ratio (SNR).
We obtain $\ln \bf=-53.5\pm1.3$ after all 20 mergers, although the injected EOS is not recovered and has only 0.04\% posterior weight.
As an increasing amount of observations is made, the TPE posterior -- expressed by the observable tidal deformability $\Lambda_{1.4}$ (middle) and radius $R_{1.4}$ (bottom) of a fiducial \SI{1.4}{M_\odot} NS -- narrows down continuously.
The combined estimate for $\Lambda_{1.4}$ decreases until it settles at $\Lambda_{1.4,\rm TPE}=204^{+4}_{-10}$ (90\% CI).
This falls just below the corresponding injection value $\Lambda_{1.4, \rm inj}=212$.
In contrast, the posterior obtained from individual runs typically overestimates the injected tidal deformability.
This relates to our conservative prior choice which prefers EOSs with high TOV masses.
As these are typically associated with higher \textLambda\ values, individual runs are biased towards overestimates of $\Lambda_{1.4}$.
The joint estimate approaches a more realistic limit only  as data from more runs and a wider range of component masses is included.

That the injection value is not recovered within 90\% CI of the joint posterior is primarily due to a systematic overestimate of the luminosity distance $d_{\rm L}$ or, equivalently, the redshift. 
The observed mass parameters are degenerate in redshift, while \textLambda\ is determined by the component masses in their source frame (i.e., not redshifted).
The overestimate in $d_{\rm L}$ leads to an underestimate of these masses.
Low masses correspond to higher deformability, an effect that the sampling algorithm will naturally compensate by selecting EOSs of more compact NSs to match the measured tidal effects.

The even larger underestimate in radius with $R_{1.4,\rm TPE}=10.80^{+0.12}_{-0.10}\,{\rm km}$ is aided by the fact that our injection EOS happens to exhibit the highest radius ($R_{1.4,\rm inj}=\SI{11.0}{km}$) among the EOSs that live in a narrow \textLambda-band around the injection value at \SI{1.4}{M_\odot}.
Since observations of the inspiral signal are not radius sensitive, we would thus expect a radius underestimate even if we had observed a more accurate \textLambda\ recovery.
This highlights the fact that the NS radius and its tidal deformability are not fully equivalent quantities.

The strong preference for the TPE Hamiltonian is explained by the fact that this low-\textLambda\ regime is only sparsely populated by the on-average stiffer \mve\ EOSs.
The high resolution of the ET effectively rules out \mve\ after a sufficient amount of signals.
After the 13th detection, only two EOSs populate more than 90\% of the \mve\ posterior space and cause the apparent jumps of the median estimate. 

Conversely, the corresponding plot for the \mve\ injection on the right of Fig. \ref{fig:main_result} demonstrates a much weaker model preference at $\ln \bf=3.8\pm1.3$. 
This is linked to the fact that the TPE model naturally provides better support for the \mve\ \textLambda\ distribution than vice versa.
The six events with the lowest SNR even seem to favor the TPE model to various degree.
A low SNR in the con\-sidered distance range is typically related to low inclination angles.
Due to the inclination-distance degeneracy, these systems are especially prone to overestimates of the luminosity distance, leading to an apparent reduction of the component masses which further benefits the softer TPE EOSs. 
Nevertheless, the \mve\ posterior is at $\Lambda_{1.4,\ve}=299^{+10}_{-7}$ in good agreement with $\Lambda_{1.4,\rm inj}=300$, while the TPE posterior overestimates it at $\Lambda_{1.4,\rm TPE}=315^{+6}_{-3}$.
\begin{figure*}[t]
    \centering
    \includegraphics[width=\textwidth]{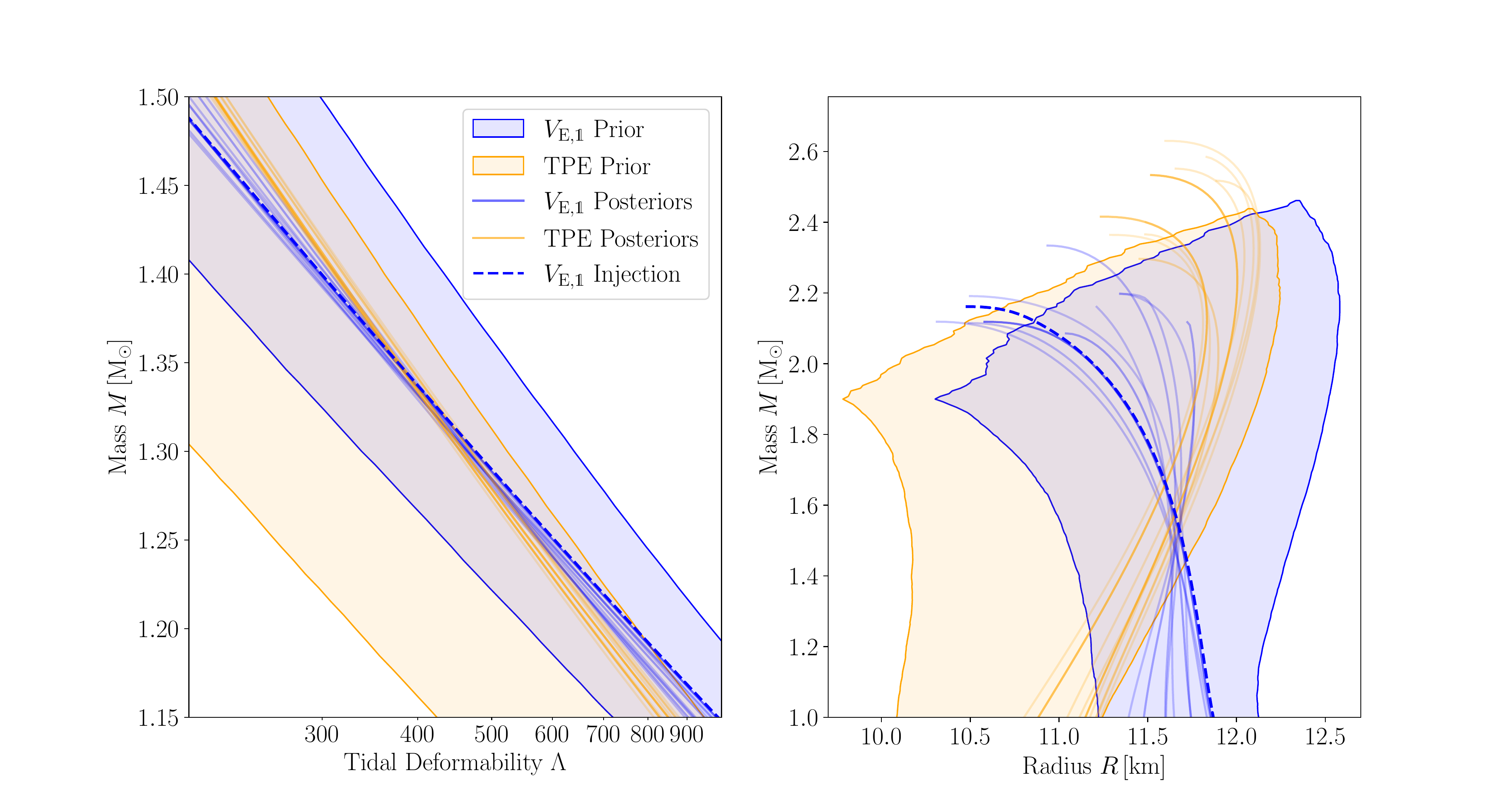}
    \caption{
    \textbf{M-\textLambda\ and M-R relation for EOS prior and posterior.} 
    Shaded bands indicate regions in prior space covered by 90\% of EOSs. 
    The two sets are well distinguishable at low masses, while the parameter spaces overlap at high masses by construction where the chiral EFT-approach breaks down. 
    The injected EOS from each set is therefore chosen based on the \textLambda-distribution at a relatively low mass of \SI{1.2}{M_\odot} as that EOS from the distribution's 50th percentile which has a TOV-mass closest to a fiducial value of \SI{2.2}{M_\odot}.
    We show the \mve\ injection as a dashed line alongside the ten dominant EOSs in each set's corresponding posterior. 
    Note how the tidal deformability is best constrained around the mean mass of the underlying mass distribution.
    }
    \label{fig:m-r-lam-prior}
\end{figure*}
Fig.~\ref{fig:m-r-lam-prior} illustrates the reason for this.
It shows the ten most likely EOSs from each set together with the \mve\ injection. 
Fainter lines correspond to subdominant posterior contributions and background contours match the priors in \textLambda\ (left) as well as radius (right).
Since merging binaries form a distinct sub-population of all NSs, our injection was guided by the mass distribution of galactic BNS systems.
We see that \textLambda\ is best recovered around that distribution's mode at \SI{1.33}{M_\odot}~\cite{Ozel:2016oaf}.
As the \mve\ set is characterized by higher tidal deformability at lowest masses, the injected \textLambda\ values are then best matched by EOSs in the TPE set that stiffen considerably and early in comparison to the full prior.
It is therefore no contradiction but indicative of the probed mass range that the joint TPE posterior overestimates $\Lambda_{1.4}$.
However, we also see in the radius plot that these EOSs mostly have TOV masses of \SI{2.4}{M_\odot} and above. 
This is in conflict with evidence for the formation of a short-lived hypermassive NS in the GW170817 merger~\cite{Rezzolla:2017aly,Margalit:2017dij,Shibata:2019ctb,Ruiz:2017due}.
Cutting these EOSs by penalizing high TOV masses in the prior could reduce the erroneous TPE preference and shift the Bayes factor in favor of the injection in a more realistic setting (see supplemental material).

{\it Conclusions - }
We have performed an ET injection study to investigate if future GW observations can help to constrain the nuclear Hamiltonian in NS matter.
For our TPE and \mve\ injection choices, constrained by chiral EFT at low densities and observations at high densities, we have found $\ln\bf=-53.5\pm1.3$ and $\ln\bf=3.8\pm1.3$, respectively. 
The strength of model preference will naturally change for more or less extreme EOSs as well as for denser EOS sets.
In neither case, we were able to recover the injected EOS.
This reflects the limited mass range expected for merging NSs, only probing an accordingly limited value range in the EOS. 
Our analysis therefore suggests that while we should remain cautious in identifying strong constraints on the global EOS from upcoming GW detections, they can be highly informative on nuclear properties that contribute to the EOS.

These outcomes need to be seen in context of computational limitations and our rather conservative approach.
We performed inference on the frequency domain above \SI{30}{Hz}. 
While tidal terms are hardly distinguishable below, this has in fact cut the most promising information on the mass parameters which contribute most information in ET around \SI{5}{Hz} to \SI{8}{Hz}~\cite{Dietrich:2020eud}.
Additionally, we have only taken a detection by ET alone into consideration.
Network operation with CE and other GW detectors as well as sky localisation by detection of EM counterparts would greatly improve inference on GW parameters, including the luminosity distance, under realistic conditions and reduce the observed bias towards the TPE model.
Moreover, upper limits on the TOV mass will be refined by further multi-messenger detections of NS mergers.
We can therefore be optimistic to achieve significantly better constraints in actual science runs (see supplemental materials).

However, even under our conservative assumptions, we found clear evidence in support of either injection, particularly when considering the systems with highest SNR.
This suggests that observations with 3rd generation GW detectors alone will be able to amass the required data to decisively distinguish nuclear Hamiltonians.
Joint detections then reduce the amount of necessary events to surpass a desired level of confidence, underlining the potential of multimessenger astronomy to inform nuclear theory.
Naturally, if the true EOS proves more extreme, our results will be more constraining than if the true EOS can be described well by either Hamiltonian.
Notwithstanding, our approach is not limited to 3N interactions but can in principle constrain other parts of the Hamiltonian, too.

{\it Acknowledgements -}
The work of S.G. and I.T. was supported by the U.S. Department of Energy, Office of Science, Office of Nuclear Physics, under contract No.~DE-AC52-06NA25396, by the U.S. Department of Energy, Office of Science, Office of Advanced Scientific Computing Research, Scientific Discovery through Advanced Computing (SciDAC) NUCLEI program, and by the Laboratory Directed Research and Development program of Los Alamos National Laboratory under project number 20230315ER. 
The work of S.G. was also supported by the Department of Energy Early Career Award Program.
The work of I.T. was also supported by the Laboratory Directed Research and Development program of Los Alamos National Laboratory under project number 20220541ECR.
This research used resources of the National Energy Research Scientific Computing Center (NERSC), a U.S. Department of Energy Office of Science User Facility located at Lawrence Berkeley National Laboratory, operated under Contract No. DE-AC02-05CH11231 using NERSC award NP-ERCAP0021027, as well as computing resources on HAWK at the High-Performance Computing Center Stuttgart (HLRS) in project GWanalysis 44189.
This material is based upon work supported by NSF's LIGO Laboratory which is a major facility fully funded by the National Science Foundation.
The authors gratefully acknowledge the Italian Istituto Nazionale di Fisica Nucleare (INFN), the French Centre National
de la Recherche Scientifique (CNRS) and the Netherlands Organization for Scientific Research (NWO), for the
construction and operation of the Virgo detector and the creation and support of the EGO consortium.

\bibliography{main.bbl}

\begin{thebibliography}{78}%
\makeatletter
\providecommand \@ifxundefined [1]{%
 \@ifx{#1\undefined}
}%
\providecommand \@ifnum [1]{%
 \ifnum #1\expandafter \@firstoftwo
 \else \expandafter \@secondoftwo
 \fi
}%
\providecommand \@ifx [1]{%
 \ifx #1\expandafter \@firstoftwo
 \else \expandafter \@secondoftwo
 \fi
}%
\providecommand \natexlab [1]{#1}%
\providecommand \enquote  [1]{``#1''}%
\providecommand \bibnamefont  [1]{#1}%
\providecommand \bibfnamefont [1]{#1}%
\providecommand \citenamefont [1]{#1}%
\providecommand \href@noop [0]{\@secondoftwo}%
\providecommand \href [0]{\begingroup \@sanitize@url \@href}%
\providecommand \@href[1]{\@@startlink{#1}\@@href}%
\providecommand \@@href[1]{\endgroup#1\@@endlink}%
\providecommand \@sanitize@url [0]{\catcode `\\12\catcode `\$12\catcode
  `\&12\catcode `\#12\catcode `\^12\catcode `\_12\catcode `\%12\relax}%
\providecommand \@@startlink[1]{}%
\providecommand \@@endlink[0]{}%
\providecommand \url  [0]{\begingroup\@sanitize@url \@url }%
\providecommand \@url [1]{\endgroup\@href {#1}{\urlprefix }}%
\providecommand \urlprefix  [0]{URL }%
\providecommand \Eprint [0]{\href }%
\providecommand \doibase [0]{https://doi.org/}%
\providecommand \selectlanguage [0]{\@gobble}%
\providecommand \bibinfo  [0]{\@secondoftwo}%
\providecommand \bibfield  [0]{\@secondoftwo}%
\providecommand \translation [1]{[#1]}%
\providecommand \BibitemOpen [0]{}%
\providecommand \bibitemStop [0]{}%
\providecommand \bibitemNoStop [0]{.\EOS\space}%
\providecommand \EOS [0]{\spacefactor3000\relax}%
\providecommand \BibitemShut  [1]{\csname bibitem#1\endcsname}%
\let\auto@bib@innerbib\@empty
\bibitem [{\citenamefont {Lattimer}\ and\ \citenamefont
  {Prakash}(2001)}]{Lattimer:2000nx}%
  \BibitemOpen
  \bibfield  {author} {\bibinfo {author} {\bibfnamefont {J.}~\bibnamefont
  {Lattimer}}\ and\ \bibinfo {author} {\bibfnamefont {M.}~\bibnamefont
  {Prakash}},\ }\bibfield  {title} {\bibinfo {title} {{Neutron star structure
  and the equation of state}},\ }\href {https://doi.org/10.1086/319702}
  {\bibfield  {journal} {\bibinfo  {journal} {Astrophys.J.}\ }\textbf {\bibinfo
  {volume} {550}},\ \bibinfo {pages} {426} (\bibinfo {year} {2001})},\ \Eprint
  {https://arxiv.org/abs/astro-ph/0002232} {arXiv:astro-ph/0002232 [astro-ph]}
  \BibitemShut {NoStop}%
\bibitem [{\citenamefont {Chamel}\ and\ \citenamefont
  {Haensel}(2008)}]{Chamel:2008ca}%
  \BibitemOpen
  \bibfield  {author} {\bibinfo {author} {\bibfnamefont {N.}~\bibnamefont
  {Chamel}}\ and\ \bibinfo {author} {\bibfnamefont {P.}~\bibnamefont
  {Haensel}},\ }\bibfield  {title} {\bibinfo {title} {{Physics of Neutron Star
  Crusts}},\ }\href {https://doi.org/10.12942/lrr-2008-10} {\bibfield
  {journal} {\bibinfo  {journal} {Living Rev. Rel.}\ }\textbf {\bibinfo
  {volume} {11}},\ \bibinfo {pages} {10} (\bibinfo {year} {2008})},\ \Eprint
  {https://arxiv.org/abs/0812.3955} {arXiv:0812.3955 [astro-ph]} \BibitemShut
  {NoStop}%
\bibitem [{\citenamefont {\"Ozel}\ and\ \citenamefont
  {Freire}(2016)}]{Ozel:2016oaf}%
  \BibitemOpen
  \bibfield  {author} {\bibinfo {author} {\bibfnamefont {F.}~\bibnamefont
  {\"Ozel}}\ and\ \bibinfo {author} {\bibfnamefont {P.}~\bibnamefont
  {Freire}},\ }\bibfield  {title} {\bibinfo {title} {{Masses, Radii, and the
  Equation of State of Neutron Stars}},\ }\href
  {https://doi.org/10.1146/annurev-astro-081915-023322} {\bibfield  {journal}
  {\bibinfo  {journal} {Ann. Rev. Astron. Astrophys.}\ }\textbf {\bibinfo
  {volume} {54}},\ \bibinfo {pages} {401} (\bibinfo {year} {2016})},\ \Eprint
  {https://arxiv.org/abs/1603.02698} {arXiv:1603.02698 [astro-ph.HE]}
  \BibitemShut {NoStop}%
\bibitem [{\citenamefont {Gandolfi}\ \emph {et~al.}(2019)\citenamefont
  {Gandolfi}, \citenamefont {Lippuner}, \citenamefont {Steiner}, \citenamefont
  {Tews}, \citenamefont {Du},\ and\ \citenamefont
  {Al-Mamun}}]{Gandolfi:2019zpj}%
  \BibitemOpen
  \bibfield  {author} {\bibinfo {author} {\bibfnamefont {S.}~\bibnamefont
  {Gandolfi}}, \bibinfo {author} {\bibfnamefont {J.}~\bibnamefont {Lippuner}},
  \bibinfo {author} {\bibfnamefont {A.~W.}\ \bibnamefont {Steiner}}, \bibinfo
  {author} {\bibfnamefont {I.}~\bibnamefont {Tews}}, \bibinfo {author}
  {\bibfnamefont {X.}~\bibnamefont {Du}},\ and\ \bibinfo {author}
  {\bibfnamefont {M.}~\bibnamefont {Al-Mamun}},\ }\bibfield  {title} {\bibinfo
  {title} {{From the microscopic to the macroscopic world: from nucleons to
  neutron stars}},\ }\href {https://doi.org/10.1088/1361-6471/ab29b3}
  {\bibfield  {journal} {\bibinfo  {journal} {J. Phys. G}\ }\textbf {\bibinfo
  {volume} {46}},\ \bibinfo {pages} {103001} (\bibinfo {year} {2019})},\
  \Eprint {https://arxiv.org/abs/1903.06730} {arXiv:1903.06730 [nucl-th]}
  \BibitemShut {NoStop}%
\bibitem [{\citenamefont {Abbott}\ \emph {et~al.}(2017)\citenamefont {Abbott}
  \emph {et~al.}}]{LIGOScientific:2017qsa}%
  \BibitemOpen
  \bibfield  {author} {\bibinfo {author} {\bibfnamefont {B.~P.}\ \bibnamefont
  {Abbott}} \emph {et~al.} (\bibinfo {collaboration} {Virgo, LIGO
  Scientific}),\ }\bibfield  {title} {\bibinfo {title} {{GW170817: Observation
  of Gravitational Waves from a Binary Neutron Star Inspiral}},\ }\href
  {https://doi.org/10.1103/PhysRevLett.119.161101} {\bibfield  {journal}
  {\bibinfo  {journal} {Phys. Rev. Lett.}\ }\textbf {\bibinfo {volume} {119}},\
  \bibinfo {pages} {161101} (\bibinfo {year} {2017})},\ \Eprint
  {https://arxiv.org/abs/1710.05832} {arXiv:1710.05832 [gr-qc]} \BibitemShut
  {NoStop}%
\bibitem [{\citenamefont {Abbott}\ \emph {et~al.}(2018)\citenamefont {Abbott}
  \emph {et~al.}}]{Abbott:2018exr}%
  \BibitemOpen
  \bibfield  {author} {\bibinfo {author} {\bibfnamefont {B.~P.}\ \bibnamefont
  {Abbott}} \emph {et~al.} (\bibinfo {collaboration} {Virgo, LIGO
  Scientific}),\ }\bibfield  {title} {\bibinfo {title} {{GW170817: Measurements
  of neutron star radii and equation of state}},\ }\href
  {https://doi.org/10.1103/PhysRevLett.121.161101} {\bibfield  {journal}
  {\bibinfo  {journal} {Phys. Rev. Lett.}\ }\textbf {\bibinfo {volume} {121}},\
  \bibinfo {pages} {161101} (\bibinfo {year} {2018})},\ \Eprint
  {https://arxiv.org/abs/1805.11581} {arXiv:1805.11581 [gr-qc]} \BibitemShut
  {NoStop}%
\bibitem [{\citenamefont {Abbott}\ \emph {et~al.}(2019)\citenamefont {Abbott}
  \emph {et~al.}}]{Abbott:2018wiz}%
  \BibitemOpen
  \bibfield  {author} {\bibinfo {author} {\bibfnamefont {B.~P.}\ \bibnamefont
  {Abbott}} \emph {et~al.} (\bibinfo {collaboration} {LIGO Scientific,
  Virgo}),\ }\bibfield  {title} {\bibinfo {title} {{Properties of the binary
  neutron star merger GW170817}},\ }\href
  {https://doi.org/10.1103/PhysRevX.9.011001} {\bibfield  {journal} {\bibinfo
  {journal} {Phys. Rev.}\ }\textbf {\bibinfo {volume} {X9}},\ \bibinfo {pages}
  {011001} (\bibinfo {year} {2019})},\ \Eprint
  {https://arxiv.org/abs/1805.11579} {arXiv:1805.11579 [gr-qc]} \BibitemShut
  {NoStop}%
\bibitem [{\citenamefont {Bauswein}\ \emph {et~al.}(2017)\citenamefont
  {Bauswein}, \citenamefont {Just}, \citenamefont {Janka},\ and\ \citenamefont
  {Stergioulas}}]{Bauswein:2017vtn}%
  \BibitemOpen
  \bibfield  {author} {\bibinfo {author} {\bibfnamefont {A.}~\bibnamefont
  {Bauswein}}, \bibinfo {author} {\bibfnamefont {O.}~\bibnamefont {Just}},
  \bibinfo {author} {\bibfnamefont {H.-T.}\ \bibnamefont {Janka}},\ and\
  \bibinfo {author} {\bibfnamefont {N.}~\bibnamefont {Stergioulas}},\
  }\bibfield  {title} {\bibinfo {title} {{Neutron-star radius constraints from
  GW170817 and future detections}},\ }\href
  {https://doi.org/10.3847/2041-8213/aa9994} {\bibfield  {journal} {\bibinfo
  {journal} {Astrophys. J.}\ }\textbf {\bibinfo {volume} {850}},\ \bibinfo
  {pages} {L34} (\bibinfo {year} {2017})},\ \Eprint
  {https://arxiv.org/abs/1710.06843} {arXiv:1710.06843 [astro-ph.HE]}
  \BibitemShut {NoStop}%
\bibitem [{\citenamefont {Radice}\ \emph {et~al.}(2018)\citenamefont {Radice},
  \citenamefont {Perego}, \citenamefont {Zappa},\ and\ \citenamefont
  {Bernuzzi}}]{Radice:2017lry}%
  \BibitemOpen
  \bibfield  {author} {\bibinfo {author} {\bibfnamefont {D.}~\bibnamefont
  {Radice}}, \bibinfo {author} {\bibfnamefont {A.}~\bibnamefont {Perego}},
  \bibinfo {author} {\bibfnamefont {F.}~\bibnamefont {Zappa}},\ and\ \bibinfo
  {author} {\bibfnamefont {S.}~\bibnamefont {Bernuzzi}},\ }\bibfield  {title}
  {\bibinfo {title} {{GW170817: Joint Constraint on the Neutron Star Equation
  of State from Multimessenger Observations}},\ }\href
  {https://doi.org/10.3847/2041-8213/aaa402} {\bibfield  {journal} {\bibinfo
  {journal} {Astrophys. J.}\ }\textbf {\bibinfo {volume} {852}},\ \bibinfo
  {pages} {L29} (\bibinfo {year} {2018})},\ \Eprint
  {https://arxiv.org/abs/1711.03647} {arXiv:1711.03647 [astro-ph.HE]}
  \BibitemShut {NoStop}%
\bibitem [{\citenamefont {Most}\ \emph {et~al.}(2018)\citenamefont {Most},
  \citenamefont {Weih}, \citenamefont {Rezzolla},\ and\ \citenamefont
  {Schaffner-Bielich}}]{Most:2018hfd}%
  \BibitemOpen
  \bibfield  {author} {\bibinfo {author} {\bibfnamefont {E.~R.}\ \bibnamefont
  {Most}}, \bibinfo {author} {\bibfnamefont {L.~R.}\ \bibnamefont {Weih}},
  \bibinfo {author} {\bibfnamefont {L.}~\bibnamefont {Rezzolla}},\ and\
  \bibinfo {author} {\bibfnamefont {J.}~\bibnamefont {Schaffner-Bielich}},\
  }\bibfield  {title} {\bibinfo {title} {{New constraints on radii and tidal
  deformabilities of neutron stars from GW170817}},\ }\href
  {https://doi.org/10.1103/PhysRevLett.120.261103} {\bibfield  {journal}
  {\bibinfo  {journal} {Phys. Rev. Lett.}\ }\textbf {\bibinfo {volume} {120}},\
  \bibinfo {pages} {261103} (\bibinfo {year} {2018})},\ \Eprint
  {https://arxiv.org/abs/1803.00549} {arXiv:1803.00549 [gr-qc]} \BibitemShut
  {NoStop}%
\bibitem [{\citenamefont {Coughlin}\ \emph {et~al.}(2019)\citenamefont
  {Coughlin}, \citenamefont {Dietrich}, \citenamefont {Margalit},\ and\
  \citenamefont {Metzger}}]{Coughlin:2018fis}%
  \BibitemOpen
  \bibfield  {author} {\bibinfo {author} {\bibfnamefont {M.~W.}\ \bibnamefont
  {Coughlin}}, \bibinfo {author} {\bibfnamefont {T.}~\bibnamefont {Dietrich}},
  \bibinfo {author} {\bibfnamefont {B.}~\bibnamefont {Margalit}},\ and\
  \bibinfo {author} {\bibfnamefont {B.~D.}\ \bibnamefont {Metzger}},\
  }\bibfield  {title} {\bibinfo {title} {{Multimessenger Bayesian parameter
  inference of a binary neutron star merger}},\ }\href
  {https://doi.org/10.1093/mnrasl/slz133} {\bibfield  {journal} {\bibinfo
  {journal} {Monthly Notices of the Royal Astronomical Society: Letters}\
  }\textbf {\bibinfo {volume} {489}},\ \bibinfo {pages} {L91} (\bibinfo {year}
  {2019})},\ \Eprint {https://arxiv.org/abs/1812.04803} {arXiv:1812.04803}
  \BibitemShut {NoStop}%
\bibitem [{\citenamefont {Dietrich}\ \emph {et~al.}(2020)\citenamefont
  {Dietrich}, \citenamefont {Coughlin}, \citenamefont {Pang}, \citenamefont
  {Bulla}, \citenamefont {Heinzel}, \citenamefont {Issa}, \citenamefont
  {Tews},\ and\ \citenamefont {Antier}}]{Dietrich:2020efo}%
  \BibitemOpen
  \bibfield  {author} {\bibinfo {author} {\bibfnamefont {T.}~\bibnamefont
  {Dietrich}}, \bibinfo {author} {\bibfnamefont {M.~W.}\ \bibnamefont
  {Coughlin}}, \bibinfo {author} {\bibfnamefont {P.~T.~H.}\ \bibnamefont
  {Pang}}, \bibinfo {author} {\bibfnamefont {M.}~\bibnamefont {Bulla}},
  \bibinfo {author} {\bibfnamefont {J.}~\bibnamefont {Heinzel}}, \bibinfo
  {author} {\bibfnamefont {L.}~\bibnamefont {Issa}}, \bibinfo {author}
  {\bibfnamefont {I.}~\bibnamefont {Tews}},\ and\ \bibinfo {author}
  {\bibfnamefont {S.}~\bibnamefont {Antier}},\ }\bibfield  {title} {\bibinfo
  {title} {{Multimessenger constraints on the neutron-star equation of state
  and the Hubble constant}},\ }\href {https://doi.org/10.1126/science.abb4317}
  {\bibfield  {journal} {\bibinfo  {journal} {Science}\ }\textbf {\bibinfo
  {volume} {370}},\ \bibinfo {pages} {1450} (\bibinfo {year} {2020})},\ \Eprint
  {https://arxiv.org/abs/2002.11355} {arXiv:2002.11355 [astro-ph.HE]}
  \BibitemShut {NoStop}%
\bibitem [{\citenamefont {Riley}\ \emph {et~al.}(2019)\citenamefont {Riley}
  \emph {et~al.}}]{Riley:2019yda}%
  \BibitemOpen
  \bibfield  {author} {\bibinfo {author} {\bibfnamefont {T.~E.}\ \bibnamefont
  {Riley}} \emph {et~al.},\ }\bibfield  {title} {\bibinfo {title} {{A $NICER$
  View of PSR J0030+0451: Millisecond Pulsar Parameter Estimation}},\ }\href
  {https://doi.org/10.3847/2041-8213/ab481c} {\bibfield  {journal} {\bibinfo
  {journal} {Astrophys. J. Lett.}\ }\textbf {\bibinfo {volume} {887}},\
  \bibinfo {pages} {L21} (\bibinfo {year} {2019})},\ \Eprint
  {https://arxiv.org/abs/1912.05702} {arXiv:1912.05702 [astro-ph.HE]}
  \BibitemShut {NoStop}%
\bibitem [{\citenamefont {Miller}\ \emph {et~al.}(2019)\citenamefont {Miller}
  \emph {et~al.}}]{Miller:2019cac}%
  \BibitemOpen
  \bibfield  {author} {\bibinfo {author} {\bibfnamefont {M.~C.}\ \bibnamefont
  {Miller}} \emph {et~al.},\ }\bibfield  {title} {\bibinfo {title} {{PSR
  J0030+0451 Mass and Radius from $NICER$ Data and Implications for the
  Properties of Neutron Star Matter}},\ }\href
  {https://doi.org/10.3847/2041-8213/ab50c5} {\bibfield  {journal} {\bibinfo
  {journal} {Astrophys. J. Lett.}\ }\textbf {\bibinfo {volume} {887}},\
  \bibinfo {pages} {L24} (\bibinfo {year} {2019})},\ \Eprint
  {https://arxiv.org/abs/1912.05705} {arXiv:1912.05705 [astro-ph.HE]}
  \BibitemShut {NoStop}%
\bibitem [{\citenamefont {Riley}\ \emph {et~al.}(2021)\citenamefont {Riley}
  \emph {et~al.}}]{Riley:2021pdl}%
  \BibitemOpen
  \bibfield  {author} {\bibinfo {author} {\bibfnamefont {T.~E.}\ \bibnamefont
  {Riley}} \emph {et~al.},\ }\bibfield  {title} {\bibinfo {title} {{A NICER
  View of the Massive Pulsar PSR J0740+6620 Informed by Radio Timing and
  XMM-Newton Spectroscopy}},\ }\href {https://doi.org/10.3847/2041-8213/ac0a81}
  {\bibfield  {journal} {\bibinfo  {journal} {Astrophys. J. Lett.}\ }\textbf
  {\bibinfo {volume} {918}},\ \bibinfo {pages} {L27} (\bibinfo {year}
  {2021})},\ \Eprint {https://arxiv.org/abs/2105.06980} {arXiv:2105.06980
  [astro-ph.HE]} \BibitemShut {NoStop}%
\bibitem [{\citenamefont {Miller}\ \emph {et~al.}(2021)\citenamefont {Miller}
  \emph {et~al.}}]{Miller:2021qha}%
  \BibitemOpen
  \bibfield  {author} {\bibinfo {author} {\bibfnamefont {M.~C.}\ \bibnamefont
  {Miller}} \emph {et~al.},\ }\bibfield  {title} {\bibinfo {title} {{The Radius
  of PSR J0740+6620 from NICER and XMM-Newton Data}},\ }\href
  {https://doi.org/10.3847/2041-8213/ac089b} {\bibfield  {journal} {\bibinfo
  {journal} {Astrophys. J. Lett.}\ }\textbf {\bibinfo {volume} {918}},\
  \bibinfo {pages} {L28} (\bibinfo {year} {2021})},\ \Eprint
  {https://arxiv.org/abs/2105.06979} {arXiv:2105.06979 [astro-ph.HE]}
  \BibitemShut {NoStop}%
\bibitem [{\citenamefont {Epelbaum}\ \emph {et~al.}(2009)\citenamefont
  {Epelbaum}, \citenamefont {Hammer},\ and\ \citenamefont
  {Meissner}}]{Epelbaum:2008ga}%
  \BibitemOpen
  \bibfield  {author} {\bibinfo {author} {\bibfnamefont {E.}~\bibnamefont
  {Epelbaum}}, \bibinfo {author} {\bibfnamefont {H.-W.}\ \bibnamefont
  {Hammer}},\ and\ \bibinfo {author} {\bibfnamefont {U.-G.}\ \bibnamefont
  {Meissner}},\ }\bibfield  {title} {\bibinfo {title} {{Modern Theory of
  Nuclear Forces}},\ }\href {https://doi.org/10.1103/RevModPhys.81.1773}
  {\bibfield  {journal} {\bibinfo  {journal} {Rev. Mod. Phys.}\ }\textbf
  {\bibinfo {volume} {81}},\ \bibinfo {pages} {1773} (\bibinfo {year}
  {2009})},\ \Eprint {https://arxiv.org/abs/0811.1338} {arXiv:0811.1338
  [nucl-th]} \BibitemShut {NoStop}%
\bibitem [{\citenamefont {Machleidt}\ and\ \citenamefont
  {Entem}(2011)}]{Machleidt:2011zz}%
  \BibitemOpen
  \bibfield  {author} {\bibinfo {author} {\bibfnamefont {R.}~\bibnamefont
  {Machleidt}}\ and\ \bibinfo {author} {\bibfnamefont {D.~R.}\ \bibnamefont
  {Entem}},\ }\bibfield  {title} {\bibinfo {title} {{Chiral effective field
  theory and nuclear forces}},\ }\href
  {https://doi.org/10.1016/j.physrep.2011.02.001} {\bibfield  {journal}
  {\bibinfo  {journal} {Phys. Rept.}\ }\textbf {\bibinfo {volume} {503}},\
  \bibinfo {pages} {1} (\bibinfo {year} {2011})},\ \Eprint
  {https://arxiv.org/abs/1105.2919} {arXiv:1105.2919 [nucl-th]} \BibitemShut
  {NoStop}%
\bibitem [{\citenamefont {Otsuka}\ \emph {et~al.}(2010)\citenamefont {Otsuka},
  \citenamefont {Suzuki}, \citenamefont {Holt}, \citenamefont {Schwenk},\ and\
  \citenamefont {Akaishi}}]{Otsuka:2009cs}%
  \BibitemOpen
  \bibfield  {author} {\bibinfo {author} {\bibfnamefont {T.}~\bibnamefont
  {Otsuka}}, \bibinfo {author} {\bibfnamefont {T.}~\bibnamefont {Suzuki}},
  \bibinfo {author} {\bibfnamefont {J.~D.}\ \bibnamefont {Holt}}, \bibinfo
  {author} {\bibfnamefont {A.}~\bibnamefont {Schwenk}},\ and\ \bibinfo {author}
  {\bibfnamefont {Y.}~\bibnamefont {Akaishi}},\ }\bibfield  {title} {\bibinfo
  {title} {{Three-body forces and the limit of oxygen isotopes}},\ }\href
  {https://doi.org/10.1103/PhysRevLett.105.032501} {\bibfield  {journal}
  {\bibinfo  {journal} {Phys. Rev. Lett.}\ }\textbf {\bibinfo {volume} {105}},\
  \bibinfo {pages} {032501} (\bibinfo {year} {2010})},\ \Eprint
  {https://arxiv.org/abs/0908.2607} {arXiv:0908.2607 [nucl-th]} \BibitemShut
  {NoStop}%
\bibitem [{\citenamefont {Wienholtz}\ \emph {et~al.}(2013)\citenamefont
  {Wienholtz} \emph {et~al.}}]{Wienholtz:2013nya}%
  \BibitemOpen
  \bibfield  {author} {\bibinfo {author} {\bibfnamefont {F.}~\bibnamefont
  {Wienholtz}} \emph {et~al.},\ }\bibfield  {title} {\bibinfo {title} {{Masses
  of exotic calcium isotopes pin down nuclear forces}},\ }\href
  {https://doi.org/10.1038/nature12226} {\bibfield  {journal} {\bibinfo
  {journal} {Nature}\ }\textbf {\bibinfo {volume} {498}},\ \bibinfo {pages}
  {346} (\bibinfo {year} {2013})}\BibitemShut {NoStop}%
\bibitem [{\citenamefont {Day}(1983)}]{Day:1983gga}%
  \BibitemOpen
  \bibfield  {author} {\bibinfo {author} {\bibfnamefont {B.~D.}\ \bibnamefont
  {Day}},\ }\bibfield  {title} {\bibinfo {title} {{Nuclear Saturation and
  Nuclear Forces}},\ }\href@noop {} {\bibfield  {journal} {\bibinfo  {journal}
  {Comments Nucl. Part. Phys.}\ }\textbf {\bibinfo {volume} {11}},\ \bibinfo
  {pages} {115} (\bibinfo {year} {1983})}\BibitemShut {NoStop}%
\bibitem [{\citenamefont {Drischler}\ \emph {et~al.}(2019)\citenamefont
  {Drischler}, \citenamefont {Hebeler},\ and\ \citenamefont
  {Schwenk}}]{Drischler:2017wtt}%
  \BibitemOpen
  \bibfield  {author} {\bibinfo {author} {\bibfnamefont {C.}~\bibnamefont
  {Drischler}}, \bibinfo {author} {\bibfnamefont {K.}~\bibnamefont {Hebeler}},\
  and\ \bibinfo {author} {\bibfnamefont {A.}~\bibnamefont {Schwenk}},\
  }\bibfield  {title} {\bibinfo {title} {{Chiral interactions up to
  next-to-next-to-next-to-leading order and nuclear saturation}},\ }\href
  {https://doi.org/10.1103/PhysRevLett.122.042501} {\bibfield  {journal}
  {\bibinfo  {journal} {Phys. Rev. Lett.}\ }\textbf {\bibinfo {volume} {122}},\
  \bibinfo {pages} {042501} (\bibinfo {year} {2019})},\ \Eprint
  {https://arxiv.org/abs/1710.08220} {arXiv:1710.08220 [nucl-th]} \BibitemShut
  {NoStop}%
\bibitem [{\citenamefont {Lonardoni}\ \emph {et~al.}(2020)\citenamefont
  {Lonardoni}, \citenamefont {Tews}, \citenamefont {Gandolfi},\ and\
  \citenamefont {Carlson}}]{Lonardoni:2020}%
  \BibitemOpen
  \bibfield  {author} {\bibinfo {author} {\bibfnamefont {D.}~\bibnamefont
  {Lonardoni}}, \bibinfo {author} {\bibfnamefont {I.}~\bibnamefont {Tews}},
  \bibinfo {author} {\bibfnamefont {S.}~\bibnamefont {Gandolfi}},\ and\
  \bibinfo {author} {\bibfnamefont {J.}~\bibnamefont {Carlson}},\ }\bibfield
  {title} {\bibinfo {title} {Nuclear and neutron-star matter from local chiral
  interactions},\ }\href {https://doi.org/10.1103/PhysRevResearch.2.022033}
  {\bibfield  {journal} {\bibinfo  {journal} {Phys. Rev. Res.}\ }\textbf
  {\bibinfo {volume} {2}},\ \bibinfo {pages} {022033} (\bibinfo {year}
  {2020})}\BibitemShut {NoStop}%
\bibitem [{\citenamefont {Navratil}\ \emph {et~al.}(2009)\citenamefont
  {Navratil}, \citenamefont {Quaglioni}, \citenamefont {Stetcu},\ and\
  \citenamefont {Barrett}}]{Navratil:2009ut}%
  \BibitemOpen
  \bibfield  {author} {\bibinfo {author} {\bibfnamefont {P.}~\bibnamefont
  {Navratil}}, \bibinfo {author} {\bibfnamefont {S.}~\bibnamefont {Quaglioni}},
  \bibinfo {author} {\bibfnamefont {I.}~\bibnamefont {Stetcu}},\ and\ \bibinfo
  {author} {\bibfnamefont {B.~R.}\ \bibnamefont {Barrett}},\ }\bibfield
  {title} {\bibinfo {title} {{Recent developments in no-core shell-model
  calculations}},\ }\href {https://doi.org/10.1088/0954-3899/36/8/083101}
  {\bibfield  {journal} {\bibinfo  {journal} {J. Phys. G}\ }\textbf {\bibinfo
  {volume} {36}},\ \bibinfo {pages} {083101} (\bibinfo {year} {2009})},\
  \Eprint {https://arxiv.org/abs/0904.0463} {arXiv:0904.0463 [nucl-th]}
  \BibitemShut {NoStop}%
\bibitem [{\citenamefont {Hebeler}\ \emph {et~al.}(2013)\citenamefont
  {Hebeler}, \citenamefont {Lattimer}, \citenamefont {Pethick},\ and\
  \citenamefont {Schwenk}}]{Hebeler:2013nza}%
  \BibitemOpen
  \bibfield  {author} {\bibinfo {author} {\bibfnamefont {K.}~\bibnamefont
  {Hebeler}}, \bibinfo {author} {\bibfnamefont {J.~M.}\ \bibnamefont
  {Lattimer}}, \bibinfo {author} {\bibfnamefont {C.~J.}\ \bibnamefont
  {Pethick}},\ and\ \bibinfo {author} {\bibfnamefont {A.}~\bibnamefont
  {Schwenk}},\ }\bibfield  {title} {\bibinfo {title} {{Equation of state and
  neutron star properties constrained by nuclear physics and observation}},\
  }\href {https://doi.org/10.1088/0004-637X/773/1/11} {\bibfield  {journal}
  {\bibinfo  {journal} {Astrophys. J.}\ }\textbf {\bibinfo {volume} {773}},\
  \bibinfo {pages} {11} (\bibinfo {year} {2013})},\ \Eprint
  {https://arxiv.org/abs/1303.4662} {arXiv:1303.4662 [astro-ph.SR]}
  \BibitemShut {NoStop}%
\bibitem [{\citenamefont {Lynn}\ \emph {et~al.}(2017)\citenamefont {Lynn},
  \citenamefont {Tews}, \citenamefont {Carlson}, \citenamefont {Gandolfi},
  \citenamefont {Gezerlis}, \citenamefont {Schmidt},\ and\ \citenamefont
  {Schwenk}}]{Lynn:2017fxg}%
  \BibitemOpen
  \bibfield  {author} {\bibinfo {author} {\bibfnamefont {J.~E.}\ \bibnamefont
  {Lynn}}, \bibinfo {author} {\bibfnamefont {I.}~\bibnamefont {Tews}}, \bibinfo
  {author} {\bibfnamefont {J.}~\bibnamefont {Carlson}}, \bibinfo {author}
  {\bibfnamefont {S.}~\bibnamefont {Gandolfi}}, \bibinfo {author}
  {\bibfnamefont {A.}~\bibnamefont {Gezerlis}}, \bibinfo {author}
  {\bibfnamefont {K.~E.}\ \bibnamefont {Schmidt}},\ and\ \bibinfo {author}
  {\bibfnamefont {A.}~\bibnamefont {Schwenk}},\ }\bibfield  {title} {\bibinfo
  {title} {{Quantum Monte Carlo calculations of light nuclei with local chiral
  two- and three-nucleon interactions}},\ }\href
  {https://doi.org/10.1103/PhysRevC.96.054007} {\bibfield  {journal} {\bibinfo
  {journal} {Phys. Rev. C}\ }\textbf {\bibinfo {volume} {96}},\ \bibinfo
  {pages} {054007} (\bibinfo {year} {2017})},\ \Eprint
  {https://arxiv.org/abs/1706.07668} {arXiv:1706.07668 [nucl-th]} \BibitemShut
  {NoStop}%
\bibitem [{\citenamefont {Punturo}\ \emph {et~al.}(2010)\citenamefont
  {Punturo}, \citenamefont {Abernathy}, \citenamefont {Acernese}, \citenamefont
  {Allen}, \citenamefont {Andersson} \emph {et~al.}}]{Punturo:2010zz}%
  \BibitemOpen
  \bibfield  {author} {\bibinfo {author} {\bibfnamefont {M.}~\bibnamefont
  {Punturo}}, \bibinfo {author} {\bibfnamefont {M.}~\bibnamefont {Abernathy}},
  \bibinfo {author} {\bibfnamefont {F.}~\bibnamefont {Acernese}}, \bibinfo
  {author} {\bibfnamefont {B.}~\bibnamefont {Allen}}, \bibinfo {author}
  {\bibfnamefont {N.}~\bibnamefont {Andersson}}, \emph {et~al.},\ }\bibfield
  {title} {\bibinfo {title} {{The Einstein Telescope: A third-generation
  gravitational wave observatory}},\ }\href
  {https://doi.org/10.1088/0264-9381/27/19/194002} {\bibfield  {journal}
  {\bibinfo  {journal} {Class.Quant.Grav.}\ }\textbf {\bibinfo {volume} {27}},\
  \bibinfo {pages} {194002} (\bibinfo {year} {2010})}\BibitemShut {NoStop}%
\bibitem [{\citenamefont {Hild}\ \emph {et~al.}(2011)\citenamefont {Hild} \emph
  {et~al.}}]{Hild:2010id}%
  \BibitemOpen
  \bibfield  {author} {\bibinfo {author} {\bibfnamefont {S.}~\bibnamefont
  {Hild}} \emph {et~al.},\ }\bibfield  {title} {\bibinfo {title} {{Sensitivity
  Studies for Third-Generation Gravitational Wave Observatories}},\ }\href
  {https://doi.org/10.1088/0264-9381/28/9/094013} {\bibfield  {journal}
  {\bibinfo  {journal} {Class. Quant. Grav.}\ }\textbf {\bibinfo {volume}
  {28}},\ \bibinfo {pages} {094013} (\bibinfo {year} {2011})},\ \Eprint
  {https://arxiv.org/abs/1012.0908} {arXiv:1012.0908 [gr-qc]} \BibitemShut
  {NoStop}%
\bibitem [{\citenamefont {Tews}\ \emph {et~al.}(2018)\citenamefont {Tews},
  \citenamefont {Margueron},\ and\ \citenamefont {Reddy}}]{Tews:2018iwm}%
  \BibitemOpen
  \bibfield  {author} {\bibinfo {author} {\bibfnamefont {I.}~\bibnamefont
  {Tews}}, \bibinfo {author} {\bibfnamefont {J.}~\bibnamefont {Margueron}},\
  and\ \bibinfo {author} {\bibfnamefont {S.}~\bibnamefont {Reddy}},\ }\bibfield
   {title} {\bibinfo {title} {{Critical examination of constraints on the
  equation of state of dense matter obtained from GW170817}},\ }\href
  {https://doi.org/10.1103/PhysRevC.98.045804} {\bibfield  {journal} {\bibinfo
  {journal} {Phys. Rev. C}\ }\textbf {\bibinfo {volume} {98}},\ \bibinfo
  {pages} {045804} (\bibinfo {year} {2018})},\ \Eprint
  {https://arxiv.org/abs/1804.02783} {arXiv:1804.02783 [nucl-th]} \BibitemShut
  {NoStop}%
\bibitem [{\citenamefont {Schmidt}\ and\ \citenamefont
  {Fantoni}(1999)}]{Schmidt:1999lik}%
  \BibitemOpen
  \bibfield  {author} {\bibinfo {author} {\bibfnamefont {K.~E.}\ \bibnamefont
  {Schmidt}}\ and\ \bibinfo {author} {\bibfnamefont {S.}~\bibnamefont
  {Fantoni}},\ }\bibfield  {title} {\bibinfo {title} {{A quantum Monte Carlo
  method for nucleon systems}},\ }\href
  {https://doi.org/10.1016/S0370-2693(98)01522-6} {\bibfield  {journal}
  {\bibinfo  {journal} {Phys. Lett. B}\ }\textbf {\bibinfo {volume} {446}},\
  \bibinfo {pages} {99} (\bibinfo {year} {1999})}\BibitemShut {NoStop}%
\bibitem [{\citenamefont {Carlson}\ \emph {et~al.}(2015)\citenamefont
  {Carlson}, \citenamefont {Gandolfi}, \citenamefont {Pederiva}, \citenamefont
  {Pieper}, \citenamefont {Schiavilla}, \citenamefont {Schmidt},\ and\
  \citenamefont {Wiringa}}]{Carlson:2014vla}%
  \BibitemOpen
  \bibfield  {author} {\bibinfo {author} {\bibfnamefont {J.}~\bibnamefont
  {Carlson}}, \bibinfo {author} {\bibfnamefont {S.}~\bibnamefont {Gandolfi}},
  \bibinfo {author} {\bibfnamefont {F.}~\bibnamefont {Pederiva}}, \bibinfo
  {author} {\bibfnamefont {S.~C.}\ \bibnamefont {Pieper}}, \bibinfo {author}
  {\bibfnamefont {R.}~\bibnamefont {Schiavilla}}, \bibinfo {author}
  {\bibfnamefont {K.~E.}\ \bibnamefont {Schmidt}},\ and\ \bibinfo {author}
  {\bibfnamefont {R.~B.}\ \bibnamefont {Wiringa}},\ }\bibfield  {title}
  {\bibinfo {title} {{Quantum Monte Carlo methods for nuclear physics}},\
  }\href {https://doi.org/10.1103/RevModPhys.87.1067} {\bibfield  {journal}
  {\bibinfo  {journal} {Rev. Mod. Phys.}\ }\textbf {\bibinfo {volume} {87}},\
  \bibinfo {pages} {1067} (\bibinfo {year} {2015})},\ \Eprint
  {https://arxiv.org/abs/1412.3081} {arXiv:1412.3081 [nucl-th]} \BibitemShut
  {NoStop}%
\bibitem [{\citenamefont {Lynn}\ \emph {et~al.}(2019)\citenamefont {Lynn},
  \citenamefont {Tews}, \citenamefont {Gandolfi},\ and\ \citenamefont
  {Lovato}}]{Lynn:2019rdt}%
  \BibitemOpen
  \bibfield  {author} {\bibinfo {author} {\bibfnamefont {J.~E.}\ \bibnamefont
  {Lynn}}, \bibinfo {author} {\bibfnamefont {I.}~\bibnamefont {Tews}}, \bibinfo
  {author} {\bibfnamefont {S.}~\bibnamefont {Gandolfi}},\ and\ \bibinfo
  {author} {\bibfnamefont {A.}~\bibnamefont {Lovato}},\ }\bibfield  {title}
  {\bibinfo {title} {{Quantum Monte Carlo Methods in Nuclear Physics: Recent
  Advances}},\ }\href {https://doi.org/10.1146/annurev-nucl-101918-023600}
  {\bibfield  {journal} {\bibinfo  {journal} {Ann. Rev. Nucl. Part. Sci.}\
  }\textbf {\bibinfo {volume} {69}},\ \bibinfo {pages} {279} (\bibinfo {year}
  {2019})},\ \Eprint {https://arxiv.org/abs/1901.04868} {arXiv:1901.04868
  [nucl-th]} \BibitemShut {NoStop}%
\bibitem [{\citenamefont {Gezerlis}\ \emph {et~al.}(2013)\citenamefont
  {Gezerlis}, \citenamefont {Tews}, \citenamefont {Epelbaum}, \citenamefont
  {Gandolfi}, \citenamefont {Hebeler}, \citenamefont {Nogga},\ and\
  \citenamefont {Schwenk}}]{Gezerlis:2013ipa}%
  \BibitemOpen
  \bibfield  {author} {\bibinfo {author} {\bibfnamefont {A.}~\bibnamefont
  {Gezerlis}}, \bibinfo {author} {\bibfnamefont {I.}~\bibnamefont {Tews}},
  \bibinfo {author} {\bibfnamefont {E.}~\bibnamefont {Epelbaum}}, \bibinfo
  {author} {\bibfnamefont {S.}~\bibnamefont {Gandolfi}}, \bibinfo {author}
  {\bibfnamefont {K.}~\bibnamefont {Hebeler}}, \bibinfo {author} {\bibfnamefont
  {A.}~\bibnamefont {Nogga}},\ and\ \bibinfo {author} {\bibfnamefont
  {A.}~\bibnamefont {Schwenk}},\ }\bibfield  {title} {\bibinfo {title}
  {{Quantum Monte Carlo Calculations with Chiral Effective Field Theory
  Interactions}},\ }\href {https://doi.org/10.1103/PhysRevLett.111.032501}
  {\bibfield  {journal} {\bibinfo  {journal} {Phys. Rev. Lett.}\ }\textbf
  {\bibinfo {volume} {111}},\ \bibinfo {pages} {032501} (\bibinfo {year}
  {2013})},\ \Eprint {https://arxiv.org/abs/1303.6243} {arXiv:1303.6243
  [nucl-th]} \BibitemShut {NoStop}%
\bibitem [{\citenamefont {Gezerlis}\ \emph {et~al.}(2014)\citenamefont
  {Gezerlis}, \citenamefont {Tews}, \citenamefont {Epelbaum}, \citenamefont
  {Freunek}, \citenamefont {Gandolfi}, \citenamefont {Hebeler}, \citenamefont
  {Nogga},\ and\ \citenamefont {Schwenk}}]{Gezerlis:2014zia}%
  \BibitemOpen
  \bibfield  {author} {\bibinfo {author} {\bibfnamefont {A.}~\bibnamefont
  {Gezerlis}}, \bibinfo {author} {\bibfnamefont {I.}~\bibnamefont {Tews}},
  \bibinfo {author} {\bibfnamefont {E.}~\bibnamefont {Epelbaum}}, \bibinfo
  {author} {\bibfnamefont {M.}~\bibnamefont {Freunek}}, \bibinfo {author}
  {\bibfnamefont {S.}~\bibnamefont {Gandolfi}}, \bibinfo {author}
  {\bibfnamefont {K.}~\bibnamefont {Hebeler}}, \bibinfo {author} {\bibfnamefont
  {A.}~\bibnamefont {Nogga}},\ and\ \bibinfo {author} {\bibfnamefont
  {A.}~\bibnamefont {Schwenk}},\ }\bibfield  {title} {\bibinfo {title} {{Local
  chiral effective field theory interactions and quantum Monte Carlo
  applications}},\ }\href {https://doi.org/10.1103/PhysRevC.90.054323}
  {\bibfield  {journal} {\bibinfo  {journal} {Phys. Rev. C}\ }\textbf {\bibinfo
  {volume} {90}},\ \bibinfo {pages} {054323} (\bibinfo {year} {2014})},\
  \Eprint {https://arxiv.org/abs/1406.0454} {arXiv:1406.0454 [nucl-th]}
  \BibitemShut {NoStop}%
\bibitem [{\citenamefont {Lynn}\ \emph {et~al.}(2016)\citenamefont {Lynn},
  \citenamefont {Tews}, \citenamefont {Carlson}, \citenamefont {Gandolfi},
  \citenamefont {Gezerlis}, \citenamefont {Schmidt},\ and\ \citenamefont
  {Schwenk}}]{Lynn:2015jua}%
  \BibitemOpen
  \bibfield  {author} {\bibinfo {author} {\bibfnamefont {J.~E.}\ \bibnamefont
  {Lynn}}, \bibinfo {author} {\bibfnamefont {I.}~\bibnamefont {Tews}}, \bibinfo
  {author} {\bibfnamefont {J.}~\bibnamefont {Carlson}}, \bibinfo {author}
  {\bibfnamefont {S.}~\bibnamefont {Gandolfi}}, \bibinfo {author}
  {\bibfnamefont {A.}~\bibnamefont {Gezerlis}}, \bibinfo {author}
  {\bibfnamefont {K.~E.}\ \bibnamefont {Schmidt}},\ and\ \bibinfo {author}
  {\bibfnamefont {A.}~\bibnamefont {Schwenk}},\ }\bibfield  {title} {\bibinfo
  {title} {{Chiral Three-Nucleon Interactions in Light Nuclei, Neutron-$\alpha$
  Scattering, and Neutron Matter}},\ }\href
  {https://doi.org/10.1103/PhysRevLett.116.062501} {\bibfield  {journal}
  {\bibinfo  {journal} {Phys. Rev. Lett.}\ }\textbf {\bibinfo {volume} {116}},\
  \bibinfo {pages} {062501} (\bibinfo {year} {2016})},\ \Eprint
  {https://arxiv.org/abs/1509.03470} {arXiv:1509.03470 [nucl-th]} \BibitemShut
  {NoStop}%
\bibitem [{Note1()}]{Note1}%
  \BibitemOpen
  \bibinfo {note} {We do not investigate the $V_{E,\tau }$ interaction of
  Ref.~\cite {Lynn:2015jua} because it leads to negative pressure in pure
  neutron matter below 2$\protect \,\rho _{\protect \rm sat}$.}\BibitemShut
  {Stop}%
\bibitem [{\citenamefont {Lonardoni}\ \emph {et~al.}(2018)\citenamefont
  {Lonardoni}, \citenamefont {Gandolfi}, \citenamefont {Lynn}, \citenamefont
  {Petrie}, \citenamefont {Carlson}, \citenamefont {Schmidt},\ and\
  \citenamefont {Schwenk}}]{Lonardoni:2018nob}%
  \BibitemOpen
  \bibfield  {author} {\bibinfo {author} {\bibfnamefont {D.}~\bibnamefont
  {Lonardoni}}, \bibinfo {author} {\bibfnamefont {S.}~\bibnamefont {Gandolfi}},
  \bibinfo {author} {\bibfnamefont {J.~E.}\ \bibnamefont {Lynn}}, \bibinfo
  {author} {\bibfnamefont {C.}~\bibnamefont {Petrie}}, \bibinfo {author}
  {\bibfnamefont {J.}~\bibnamefont {Carlson}}, \bibinfo {author} {\bibfnamefont
  {K.~E.}\ \bibnamefont {Schmidt}},\ and\ \bibinfo {author} {\bibfnamefont
  {A.}~\bibnamefont {Schwenk}},\ }\bibfield  {title} {\bibinfo {title}
  {{Auxiliary field diffusion Monte Carlo calculations of light and medium-mass
  nuclei with local chiral interactions}},\ }\href
  {https://doi.org/10.1103/PhysRevC.97.044318} {\bibfield  {journal} {\bibinfo
  {journal} {Phys. Rev. C}\ }\textbf {\bibinfo {volume} {97}},\ \bibinfo
  {pages} {044318} (\bibinfo {year} {2018})},\ \Eprint
  {https://arxiv.org/abs/1802.08932} {arXiv:1802.08932 [nucl-th]} \BibitemShut
  {NoStop}%
\bibitem [{\citenamefont {Hebeler}\ and\ \citenamefont
  {Schwenk}(2010)}]{Hebeler:2009iv}%
  \BibitemOpen
  \bibfield  {author} {\bibinfo {author} {\bibfnamefont {K.}~\bibnamefont
  {Hebeler}}\ and\ \bibinfo {author} {\bibfnamefont {A.}~\bibnamefont
  {Schwenk}},\ }\bibfield  {title} {\bibinfo {title} {{Chiral three-nucleon
  forces and neutron matter}},\ }\href
  {https://doi.org/10.1103/PhysRevC.82.014314} {\bibfield  {journal} {\bibinfo
  {journal} {Phys. Rev. C}\ }\textbf {\bibinfo {volume} {82}},\ \bibinfo
  {pages} {014314} (\bibinfo {year} {2010})},\ \Eprint
  {https://arxiv.org/abs/0911.0483} {arXiv:0911.0483 [nucl-th]} \BibitemShut
  {NoStop}%
\bibitem [{\citenamefont {Huth}\ \emph {et~al.}(2017)\citenamefont {Huth},
  \citenamefont {Tews}, \citenamefont {Lynn},\ and\ \citenamefont
  {Schwenk}}]{Huth:2017wzw}%
  \BibitemOpen
  \bibfield  {author} {\bibinfo {author} {\bibfnamefont {L.}~\bibnamefont
  {Huth}}, \bibinfo {author} {\bibfnamefont {I.}~\bibnamefont {Tews}}, \bibinfo
  {author} {\bibfnamefont {J.~E.}\ \bibnamefont {Lynn}},\ and\ \bibinfo
  {author} {\bibfnamefont {A.}~\bibnamefont {Schwenk}},\ }\bibfield  {title}
  {\bibinfo {title} {{Analyzing the Fierz Rearrangement Freedom for Local
  Chiral Two-Nucleon Potentials}},\ }\href
  {https://doi.org/10.1103/PhysRevC.96.054003} {\bibfield  {journal} {\bibinfo
  {journal} {Phys. Rev. C}\ }\textbf {\bibinfo {volume} {96}},\ \bibinfo
  {pages} {054003} (\bibinfo {year} {2017})},\ \Eprint
  {https://arxiv.org/abs/1708.03194} {arXiv:1708.03194 [nucl-th]} \BibitemShut
  {NoStop}%
\bibitem [{Note2()}]{Note2}%
  \BibitemOpen
  \bibinfo {note} {In principle, these regulator artifacts can be thought of as
  sub-leading 3N contact interactions, appearing first at N$^4$LO in chiral
  EFT.}\BibitemShut {Stop}%
\bibitem [{\citenamefont {Maselli}\ \emph {et~al.}(2021)\citenamefont
  {Maselli}, \citenamefont {Sabatucci},\ and\ \citenamefont
  {Benhar}}]{Maselli:2020uol}%
  \BibitemOpen
  \bibfield  {author} {\bibinfo {author} {\bibfnamefont {A.}~\bibnamefont
  {Maselli}}, \bibinfo {author} {\bibfnamefont {A.}~\bibnamefont {Sabatucci}},\
  and\ \bibinfo {author} {\bibfnamefont {O.}~\bibnamefont {Benhar}},\
  }\bibfield  {title} {\bibinfo {title} {{Constraining three-nucleon forces
  with multimessenger data}},\ }\href
  {https://doi.org/10.1103/PhysRevC.103.065804} {\bibfield  {journal} {\bibinfo
   {journal} {Phys. Rev. C}\ }\textbf {\bibinfo {volume} {103}},\ \bibinfo
  {pages} {065804} (\bibinfo {year} {2021})},\ \Eprint
  {https://arxiv.org/abs/2010.03581} {arXiv:2010.03581 [astro-ph.HE]}
  \BibitemShut {NoStop}%
\bibitem [{\citenamefont {Sabatucci}\ \emph {et~al.}(2022)\citenamefont
  {Sabatucci}, \citenamefont {Benhar}, \citenamefont {Maselli},\ and\
  \citenamefont {Pacilio}}]{Sabatucci:2022qyi}%
  \BibitemOpen
  \bibfield  {author} {\bibinfo {author} {\bibfnamefont {A.}~\bibnamefont
  {Sabatucci}}, \bibinfo {author} {\bibfnamefont {O.}~\bibnamefont {Benhar}},
  \bibinfo {author} {\bibfnamefont {A.}~\bibnamefont {Maselli}},\ and\ \bibinfo
  {author} {\bibfnamefont {C.}~\bibnamefont {Pacilio}},\ }\bibfield  {title}
  {\bibinfo {title} {{Sensitivity of neutron star observations to three-nucleon
  forces}},\ }\href {https://doi.org/10.1103/PhysRevD.106.083010} {\bibfield
  {journal} {\bibinfo  {journal} {Phys. Rev. D}\ }\textbf {\bibinfo {volume}
  {106}},\ \bibinfo {pages} {083010} (\bibinfo {year} {2022})},\ \Eprint
  {https://arxiv.org/abs/2206.11286} {arXiv:2206.11286 [astro-ph.HE]}
  \BibitemShut {NoStop}%
\bibitem [{\citenamefont {Hinderer}(2008)}]{Hinderer:2007mb}%
  \BibitemOpen
  \bibfield  {author} {\bibinfo {author} {\bibfnamefont {T.}~\bibnamefont
  {Hinderer}},\ }\bibfield  {title} {\bibinfo {title} {{Tidal Love numbers of
  neutron stars}},\ }\href {https://doi.org/10.1086/533487} {\bibfield
  {journal} {\bibinfo  {journal} {Astrophys. J.}\ }\textbf {\bibinfo {volume}
  {677}},\ \bibinfo {pages} {1216} (\bibinfo {year} {2008})},\ \Eprint
  {https://arxiv.org/abs/0711.2420} {arXiv:0711.2420 [astro-ph]} \BibitemShut
  {NoStop}%
\bibitem [{\citenamefont {Chatziioannou}(2020)}]{Chatziioannou:2020pqz}%
  \BibitemOpen
  \bibfield  {author} {\bibinfo {author} {\bibfnamefont {K.}~\bibnamefont
  {Chatziioannou}},\ }\bibfield  {title} {\bibinfo {title} {{Neutron star tidal
  deformability and equation of state constraints}},\ }\href
  {https://doi.org/10.1007/s10714-020-02754-3} {\bibfield  {journal} {\bibinfo
  {journal} {Gen. Rel. Grav.}\ }\textbf {\bibinfo {volume} {52}},\ \bibinfo
  {pages} {109} (\bibinfo {year} {2020})},\ \Eprint
  {https://arxiv.org/abs/2006.03168} {arXiv:2006.03168 [gr-qc]} \BibitemShut
  {NoStop}%
\bibitem [{\citenamefont {Skilling}(2006)}]{Skilling:2006gxv}%
  \BibitemOpen
  \bibfield  {author} {\bibinfo {author} {\bibfnamefont {J.}~\bibnamefont
  {Skilling}},\ }\bibfield  {title} {\bibinfo {title} {{Nested sampling for
  general Bayesian computation}},\ }\href {https://doi.org/10.1214/06-BA127}
  {\bibfield  {journal} {\bibinfo  {journal} {Bayesian Analysis}\ }\textbf
  {\bibinfo {volume} {1}},\ \bibinfo {pages} {833} (\bibinfo {year}
  {2006})}\BibitemShut {NoStop}%
\bibitem [{\citenamefont {Thrane}\ and\ \citenamefont
  {Talbot}(2019)}]{Thrane:2018qnx}%
  \BibitemOpen
  \bibfield  {author} {\bibinfo {author} {\bibfnamefont {E.}~\bibnamefont
  {Thrane}}\ and\ \bibinfo {author} {\bibfnamefont {C.}~\bibnamefont
  {Talbot}},\ }\bibfield  {title} {\bibinfo {title} {{An introduction to
  Bayesian inference in gravitational-wave astronomy: parameter estimation,
  model selection, and hierarchical models}},\ }\href
  {https://doi.org/10.1017/pasa.2019.2} {\bibfield  {journal} {\bibinfo
  {journal} {Publ. Astron. Soc. Austral.}\ }\textbf {\bibinfo {volume} {36}},\
  \bibinfo {pages} {e010} (\bibinfo {year} {2019})},\ \bibinfo {note}
  {[Erratum: Publ.Astron.Soc.Austral. 37, e036 (2020)]},\ \Eprint
  {https://arxiv.org/abs/1809.02293} {arXiv:1809.02293 [astro-ph.IM]}
  \BibitemShut {NoStop}%
\bibitem [{\citenamefont {Antoniadis}\ \emph {et~al.}(2013)\citenamefont
  {Antoniadis}, \citenamefont {Freire}, \citenamefont {Wex}, \citenamefont
  {Tauris}, \citenamefont {Lynch} \emph {et~al.}}]{Antoniadis:2013pzd}%
  \BibitemOpen
  \bibfield  {author} {\bibinfo {author} {\bibfnamefont {J.}~\bibnamefont
  {Antoniadis}}, \bibinfo {author} {\bibfnamefont {P.~C.}\ \bibnamefont
  {Freire}}, \bibinfo {author} {\bibfnamefont {N.}~\bibnamefont {Wex}},
  \bibinfo {author} {\bibfnamefont {T.~M.}\ \bibnamefont {Tauris}}, \bibinfo
  {author} {\bibfnamefont {R.~S.}\ \bibnamefont {Lynch}}, \emph {et~al.},\
  }\bibfield  {title} {\bibinfo {title} {{A Massive Pulsar in a Compact
  Relativistic Binary}},\ }\href {https://doi.org/10.1126/science.1233232}
  {\bibfield  {journal} {\bibinfo  {journal} {Science}\ }\textbf {\bibinfo
  {volume} {340}},\ \bibinfo {pages} {6131} (\bibinfo {year} {2013})},\ \Eprint
  {https://arxiv.org/abs/1304.6875} {arXiv:1304.6875 [astro-ph.HE]}
  \BibitemShut {NoStop}%
\bibitem [{\citenamefont {Arzoumanian}\ \emph {et~al.}(2018)\citenamefont
  {Arzoumanian} \emph {et~al.}}]{Arzoumanian:2017puf}%
  \BibitemOpen
  \bibfield  {author} {\bibinfo {author} {\bibfnamefont {Z.}~\bibnamefont
  {Arzoumanian}} \emph {et~al.} (\bibinfo {collaboration} {NANOGrav}),\
  }\bibfield  {title} {\bibinfo {title} {{The NANOGrav 11-year Data Set:
  High-precision timing of 45 Millisecond Pulsars}},\ }\href
  {https://doi.org/10.3847/1538-4365/aab5b0} {\bibfield  {journal} {\bibinfo
  {journal} {Astrophys. J. Suppl.}\ }\textbf {\bibinfo {volume} {235}},\
  \bibinfo {pages} {37} (\bibinfo {year} {2018})},\ \Eprint
  {https://arxiv.org/abs/1801.01837} {arXiv:1801.01837 [astro-ph.HE]}
  \BibitemShut {NoStop}%
\bibitem [{\citenamefont {Fonseca}\ \emph {et~al.}(2021)\citenamefont {Fonseca}
  \emph {et~al.}}]{Fonseca:2021wxt}%
  \BibitemOpen
  \bibfield  {author} {\bibinfo {author} {\bibfnamefont {E.}~\bibnamefont
  {Fonseca}} \emph {et~al.},\ }\bibfield  {title} {\bibinfo {title} {{Refined
  Mass and Geometric Measurements of the High-mass PSR J0740+6620}},\ }\href
  {https://doi.org/10.3847/2041-8213/ac03b8} {\bibfield  {journal} {\bibinfo
  {journal} {Astrophys. J. Lett.}\ }\textbf {\bibinfo {volume} {915}},\
  \bibinfo {pages} {L12} (\bibinfo {year} {2021})},\ \Eprint
  {https://arxiv.org/abs/2104.00880} {arXiv:2104.00880 [astro-ph.HE]}
  \BibitemShut {NoStop}%
\bibitem [{\citenamefont {Acernese}\ \emph {et~al.}(2015)\citenamefont
  {Acernese} \emph {et~al.}}]{VIRGO:2014yos}%
  \BibitemOpen
  \bibfield  {author} {\bibinfo {author} {\bibfnamefont {F.}~\bibnamefont
  {Acernese}} \emph {et~al.} (\bibinfo {collaboration} {VIRGO}),\ }\bibfield
  {title} {\bibinfo {title} {{Advanced Virgo: a second-generation
  interferometric gravitational wave detector}},\ }\href
  {https://doi.org/10.1088/0264-9381/32/2/024001} {\bibfield  {journal}
  {\bibinfo  {journal} {Class. Quant. Grav.}\ }\textbf {\bibinfo {volume}
  {32}},\ \bibinfo {pages} {024001} (\bibinfo {year} {2015})},\ \Eprint
  {https://arxiv.org/abs/1408.3978} {arXiv:1408.3978 [gr-qc]} \BibitemShut
  {NoStop}%
\bibitem [{\citenamefont {Aasi}\ \emph {et~al.}(2015)\citenamefont {Aasi} \emph
  {et~al.}}]{LIGOScientific:2014jea}%
  \BibitemOpen
  \bibfield  {author} {\bibinfo {author} {\bibfnamefont {J.}~\bibnamefont
  {Aasi}} \emph {et~al.} (\bibinfo {collaboration} {LIGO Scientific}),\
  }\bibfield  {title} {\bibinfo {title} {{Advanced LIGO}},\ }\href
  {https://doi.org/10.1088/0264-9381/32/7/074001} {\bibfield  {journal}
  {\bibinfo  {journal} {Class. Quant. Grav.}\ }\textbf {\bibinfo {volume}
  {32}},\ \bibinfo {pages} {074001} (\bibinfo {year} {2015})},\ \Eprint
  {https://arxiv.org/abs/1411.4547} {arXiv:1411.4547 [gr-qc]} \BibitemShut
  {NoStop}%
\bibitem [{\citenamefont {Abbott}\ \emph
  {et~al.}(2021{\natexlab{a}})\citenamefont {Abbott} \emph
  {et~al.}}]{LIGOScientific:2021djp}%
  \BibitemOpen
  \bibfield  {author} {\bibinfo {author} {\bibfnamefont {R.}~\bibnamefont
  {Abbott}} \emph {et~al.} (\bibinfo {collaboration} {LIGO Scientific, VIRGO,
  KAGRA}),\ }\bibfield  {title} {\bibinfo {title} {{GWTC-3: Compact Binary
  Coalescences Observed by LIGO and Virgo During the Second Part of the Third
  Observing Run}},\ }\href@noop {} {\bibfield  {journal} {\bibinfo  {journal}
  {arXiv e-prints}\ } (\bibinfo {year} {2021}{\natexlab{a}})},\ \Eprint
  {https://arxiv.org/abs/2111.03606} {arXiv:2111.03606 [gr-qc]} \BibitemShut
  {NoStop}%
\bibitem [{\citenamefont {Abbott}\ \emph
  {et~al.}(2021{\natexlab{b}})\citenamefont {Abbott} \emph
  {et~al.}}]{LIGOScientific:2021usb}%
  \BibitemOpen
  \bibfield  {author} {\bibinfo {author} {\bibfnamefont {R.}~\bibnamefont
  {Abbott}} \emph {et~al.} (\bibinfo {collaboration} {LIGO Scientific,
  VIRGO}),\ }\bibfield  {title} {\bibinfo {title} {{GWTC-2.1: Deep Extended
  Catalog of Compact Binary Coalescences Observed by LIGO and Virgo During the
  First Half of the Third Observing Run}},\ }\href@noop {} {\bibfield
  {journal} {\bibinfo  {journal} {arXiv e-prints}\ } (\bibinfo {year}
  {2021}{\natexlab{b}})},\ \Eprint {https://arxiv.org/abs/2108.01045}
  {arXiv:2108.01045 [gr-qc]} \BibitemShut {NoStop}%
\bibitem [{\citenamefont {Petrov}\ \emph {et~al.}(2022)\citenamefont {Petrov},
  \citenamefont {Singer}, \citenamefont {Coughlin}, \citenamefont {Kumar},
  \citenamefont {Almualla}, \citenamefont {Anand}, \citenamefont {Bulla},
  \citenamefont {Dietrich}, \citenamefont {Foucart},\ and\ \citenamefont
  {Guessoum}}]{Petrov:2021bqm}%
  \BibitemOpen
  \bibfield  {author} {\bibinfo {author} {\bibfnamefont {P.}~\bibnamefont
  {Petrov}}, \bibinfo {author} {\bibfnamefont {L.~P.}\ \bibnamefont {Singer}},
  \bibinfo {author} {\bibfnamefont {M.~W.}\ \bibnamefont {Coughlin}}, \bibinfo
  {author} {\bibfnamefont {V.}~\bibnamefont {Kumar}}, \bibinfo {author}
  {\bibfnamefont {M.}~\bibnamefont {Almualla}}, \bibinfo {author}
  {\bibfnamefont {S.}~\bibnamefont {Anand}}, \bibinfo {author} {\bibfnamefont
  {M.}~\bibnamefont {Bulla}}, \bibinfo {author} {\bibfnamefont
  {T.}~\bibnamefont {Dietrich}}, \bibinfo {author} {\bibfnamefont
  {F.}~\bibnamefont {Foucart}},\ and\ \bibinfo {author} {\bibfnamefont
  {N.}~\bibnamefont {Guessoum}},\ }\bibfield  {title} {\bibinfo {title}
  {{Data-driven Expectations for Electromagnetic Counterpart Searches Based on
  LIGO/Virgo Public Alerts}},\ }\href
  {https://doi.org/10.3847/1538-4357/ac366d} {\bibfield  {journal} {\bibinfo
  {journal} {Astrophys. J.}\ }\textbf {\bibinfo {volume} {924}},\ \bibinfo
  {pages} {54} (\bibinfo {year} {2022})},\ \Eprint
  {https://arxiv.org/abs/2108.07277} {arXiv:2108.07277 [astro-ph.HE]}
  \BibitemShut {NoStop}%
\bibitem [{\citenamefont {Colombo}\ \emph {et~al.}(2022)\citenamefont
  {Colombo}, \citenamefont {Salafia}, \citenamefont {Gabrielli}, \citenamefont
  {Ghirlanda}, \citenamefont {Giacomazzo}, \citenamefont {Perego},\ and\
  \citenamefont {Colpi}}]{Colombo:2022zzp}%
  \BibitemOpen
  \bibfield  {author} {\bibinfo {author} {\bibfnamefont {A.}~\bibnamefont
  {Colombo}}, \bibinfo {author} {\bibfnamefont {O.~S.}\ \bibnamefont
  {Salafia}}, \bibinfo {author} {\bibfnamefont {F.}~\bibnamefont {Gabrielli}},
  \bibinfo {author} {\bibfnamefont {G.}~\bibnamefont {Ghirlanda}}, \bibinfo
  {author} {\bibfnamefont {B.}~\bibnamefont {Giacomazzo}}, \bibinfo {author}
  {\bibfnamefont {A.}~\bibnamefont {Perego}},\ and\ \bibinfo {author}
  {\bibfnamefont {M.}~\bibnamefont {Colpi}},\ }\bibfield  {title} {\bibinfo
  {title} {{Multi-messenger Observations of Binary Neutron Star Mergers in the
  O4 Run}},\ }\href {https://doi.org/10.3847/1538-4357/ac8d00} {\bibfield
  {journal} {\bibinfo  {journal} {Astrophys. J.}\ }\textbf {\bibinfo {volume}
  {937}},\ \bibinfo {pages} {79} (\bibinfo {year} {2022})},\ \Eprint
  {https://arxiv.org/abs/2204.07592} {arXiv:2204.07592 [astro-ph.HE]}
  \BibitemShut {NoStop}%
\bibitem [{\citenamefont {Pacilio}\ \emph {et~al.}(2022)\citenamefont
  {Pacilio}, \citenamefont {Maselli}, \citenamefont {Fasano},\ and\
  \citenamefont {Pani}}]{Pacilio:2021}%
  \BibitemOpen
  \bibfield  {author} {\bibinfo {author} {\bibfnamefont {C.}~\bibnamefont
  {Pacilio}}, \bibinfo {author} {\bibfnamefont {A.}~\bibnamefont {Maselli}},
  \bibinfo {author} {\bibfnamefont {M.}~\bibnamefont {Fasano}},\ and\ \bibinfo
  {author} {\bibfnamefont {P.}~\bibnamefont {Pani}},\ }\bibfield  {title}
  {\bibinfo {title} {{Ranking Love Numbers for the Neutron Star Equation of
  State: The Need for Third-Generation Detectors}},\ }\href
  {https://doi.org/10.1103/PhysRevLett.128.101101} {\bibfield  {journal}
  {\bibinfo  {journal} {Phys. Rev. Lett.}\ }\textbf {\bibinfo {volume} {128}},\
  \bibinfo {pages} {101101} (\bibinfo {year} {2022})},\ \Eprint
  {https://arxiv.org/abs/2104.10035} {arXiv:2104.10035 [gr-qc]} \BibitemShut
  {NoStop}%
\bibitem [{\citenamefont {Reitze}\ \emph {et~al.}(2019)\citenamefont {Reitze}
  \emph {et~al.}}]{Reitze:2019iox}%
  \BibitemOpen
  \bibfield  {author} {\bibinfo {author} {\bibfnamefont {D.}~\bibnamefont
  {Reitze}} \emph {et~al.},\ }\bibfield  {title} {\bibinfo {title} {{Cosmic
  Explorer: The U.S. Contribution to Gravitational-Wave Astronomy beyond
  LIGO}},\ }\href@noop {} {\bibfield  {journal} {\bibinfo  {journal} {Bull. Am.
  Astron. Soc.}\ }\textbf {\bibinfo {volume} {51}},\ \bibinfo {pages} {035}
  (\bibinfo {year} {2019})},\ \Eprint {https://arxiv.org/abs/1907.04833}
  {arXiv:1907.04833 [astro-ph.IM]} \BibitemShut {NoStop}%
\bibitem [{\citenamefont {Rezzolla}\ \emph {et~al.}(2018)\citenamefont
  {Rezzolla}, \citenamefont {Most},\ and\ \citenamefont
  {Weih}}]{Rezzolla:2017aly}%
  \BibitemOpen
  \bibfield  {author} {\bibinfo {author} {\bibfnamefont {L.}~\bibnamefont
  {Rezzolla}}, \bibinfo {author} {\bibfnamefont {E.~R.}\ \bibnamefont {Most}},\
  and\ \bibinfo {author} {\bibfnamefont {L.~R.}\ \bibnamefont {Weih}},\
  }\bibfield  {title} {\bibinfo {title} {{Using gravitational-wave observations
  and quasi-universal relations to constrain the maximum mass of neutron
  stars}},\ }\href {https://doi.org/10.3847/2041-8213/aaa401} {\bibfield
  {journal} {\bibinfo  {journal} {Astrophys. J.}\ }\textbf {\bibinfo {volume}
  {852}},\ \bibinfo {pages} {L25} (\bibinfo {year} {2018})},\ \Eprint
  {https://arxiv.org/abs/1711.00314} {arXiv:1711.00314 [astro-ph.HE]}
  \BibitemShut {NoStop}%
\bibitem [{\citenamefont {Margalit}\ and\ \citenamefont
  {Metzger}(2017)}]{Margalit:2017dij}%
  \BibitemOpen
  \bibfield  {author} {\bibinfo {author} {\bibfnamefont {B.}~\bibnamefont
  {Margalit}}\ and\ \bibinfo {author} {\bibfnamefont {B.~D.}\ \bibnamefont
  {Metzger}},\ }\bibfield  {title} {\bibinfo {title} {{Constraining the Maximum
  Mass of Neutron Stars From Multi-Messenger Observations of GW170817}},\
  }\href {https://doi.org/10.3847/2041-8213/aa991c} {\bibfield  {journal}
  {\bibinfo  {journal} {Astrophys. J. Lett.}\ }\textbf {\bibinfo {volume}
  {850}},\ \bibinfo {pages} {L19} (\bibinfo {year} {2017})},\ \Eprint
  {https://arxiv.org/abs/1710.05938} {arXiv:1710.05938 [astro-ph.HE]}
  \BibitemShut {NoStop}%
\bibitem [{\citenamefont {Shibata}\ \emph {et~al.}(2019)\citenamefont
  {Shibata}, \citenamefont {Zhou}, \citenamefont {Kiuchi},\ and\ \citenamefont
  {Fujibayashi}}]{Shibata:2019ctb}%
  \BibitemOpen
  \bibfield  {author} {\bibinfo {author} {\bibfnamefont {M.}~\bibnamefont
  {Shibata}}, \bibinfo {author} {\bibfnamefont {E.}~\bibnamefont {Zhou}},
  \bibinfo {author} {\bibfnamefont {K.}~\bibnamefont {Kiuchi}},\ and\ \bibinfo
  {author} {\bibfnamefont {S.}~\bibnamefont {Fujibayashi}},\ }\bibfield
  {title} {\bibinfo {title} {{Constraint on the maximum mass of neutron stars
  using GW170817 event}},\ }\href {https://doi.org/10.1103/PhysRevD.100.023015}
  {\bibfield  {journal} {\bibinfo  {journal} {Phys. Rev.}\ }\textbf {\bibinfo
  {volume} {D100}},\ \bibinfo {pages} {023015} (\bibinfo {year} {2019})},\
  \Eprint {https://arxiv.org/abs/1905.03656} {arXiv:1905.03656 [astro-ph.HE]}
  \BibitemShut {NoStop}%
\bibitem [{\citenamefont {Ruiz}\ \emph {et~al.}(2018)\citenamefont {Ruiz},
  \citenamefont {Shapiro},\ and\ \citenamefont {Tsokaros}}]{Ruiz:2017due}%
  \BibitemOpen
  \bibfield  {author} {\bibinfo {author} {\bibfnamefont {M.}~\bibnamefont
  {Ruiz}}, \bibinfo {author} {\bibfnamefont {S.~L.}\ \bibnamefont {Shapiro}},\
  and\ \bibinfo {author} {\bibfnamefont {A.}~\bibnamefont {Tsokaros}},\
  }\bibfield  {title} {\bibinfo {title} {{GW170817, General Relativistic
  Magnetohydrodynamic Simulations, and the Neutron Star Maximum Mass}},\ }\href
  {https://doi.org/10.1103/PhysRevD.97.021501} {\bibfield  {journal} {\bibinfo
  {journal} {Phys. Rev.}\ }\textbf {\bibinfo {volume} {D97}},\ \bibinfo {pages}
  {021501} (\bibinfo {year} {2018})},\ \Eprint
  {https://arxiv.org/abs/1711.00473} {arXiv:1711.00473 [astro-ph.HE]}
  \BibitemShut {NoStop}%
\bibitem [{\citenamefont {Dietrich}\ \emph {et~al.}(2021)\citenamefont
  {Dietrich}, \citenamefont {Hinderer},\ and\ \citenamefont
  {Samajdar}}]{Dietrich:2020eud}%
  \BibitemOpen
  \bibfield  {author} {\bibinfo {author} {\bibfnamefont {T.}~\bibnamefont
  {Dietrich}}, \bibinfo {author} {\bibfnamefont {T.}~\bibnamefont {Hinderer}},\
  and\ \bibinfo {author} {\bibfnamefont {A.}~\bibnamefont {Samajdar}},\
  }\bibfield  {title} {\bibinfo {title} {{Interpreting Binary Neutron Star
  Mergers: Describing the Binary Neutron Star Dynamics, Modelling Gravitational
  Waveforms, and Analyzing Detections}},\ }\href
  {https://doi.org/10.1007/s10714-020-02751-6} {\bibfield  {journal} {\bibinfo
  {journal} {Gen. Rel. Grav.}\ }\textbf {\bibinfo {volume} {53}},\ \bibinfo
  {pages} {27} (\bibinfo {year} {2021})},\ \Eprint
  {https://arxiv.org/abs/2004.02527} {arXiv:2004.02527 [gr-qc]} \BibitemShut
  {NoStop}%
\bibitem [{\citenamefont {Dietrich}\ \emph {et~al.}(2019)\citenamefont
  {Dietrich}, \citenamefont {Samajdar}, \citenamefont {Khan}, \citenamefont
  {Johnson-McDaniel}, \citenamefont {Dudi},\ and\ \citenamefont
  {Tichy}}]{Dietrich:2019kaq}%
  \BibitemOpen
  \bibfield  {author} {\bibinfo {author} {\bibfnamefont {T.}~\bibnamefont
  {Dietrich}}, \bibinfo {author} {\bibfnamefont {A.}~\bibnamefont {Samajdar}},
  \bibinfo {author} {\bibfnamefont {S.}~\bibnamefont {Khan}}, \bibinfo {author}
  {\bibfnamefont {N.~K.}\ \bibnamefont {Johnson-McDaniel}}, \bibinfo {author}
  {\bibfnamefont {R.}~\bibnamefont {Dudi}},\ and\ \bibinfo {author}
  {\bibfnamefont {W.}~\bibnamefont {Tichy}},\ }\bibfield  {title} {\bibinfo
  {title} {{Improving the NRTidal model for binary neutron star systems}},\
  }\href {https://doi.org/10.1103/PhysRevD.100.044003} {\bibfield  {journal}
  {\bibinfo  {journal} {Phys. Rev.}\ }\textbf {\bibinfo {volume} {D100}},\
  \bibinfo {pages} {044003} (\bibinfo {year} {2019})},\ \Eprint
  {https://arxiv.org/abs/1905.06011} {arXiv:1905.06011 [gr-qc]} \BibitemShut
  {NoStop}%
\bibitem [{\citenamefont {Martinez}\ \emph {et~al.}(2015)\citenamefont
  {Martinez}, \citenamefont {Stovall}, \citenamefont {Freire}, \citenamefont
  {Deneva}, \citenamefont {Jenet}, \citenamefont {McLaughlin}, \citenamefont
  {Bagchi}, \citenamefont {Bates},\ and\ \citenamefont
  {Ridolfi}}]{Martinez:2015mya}%
  \BibitemOpen
  \bibfield  {author} {\bibinfo {author} {\bibfnamefont {J.~G.}\ \bibnamefont
  {Martinez}}, \bibinfo {author} {\bibfnamefont {K.}~\bibnamefont {Stovall}},
  \bibinfo {author} {\bibfnamefont {P.~C.~C.}\ \bibnamefont {Freire}}, \bibinfo
  {author} {\bibfnamefont {J.~S.}\ \bibnamefont {Deneva}}, \bibinfo {author}
  {\bibfnamefont {F.~A.}\ \bibnamefont {Jenet}}, \bibinfo {author}
  {\bibfnamefont {M.~A.}\ \bibnamefont {McLaughlin}}, \bibinfo {author}
  {\bibfnamefont {M.}~\bibnamefont {Bagchi}}, \bibinfo {author} {\bibfnamefont
  {S.~D.}\ \bibnamefont {Bates}},\ and\ \bibinfo {author} {\bibfnamefont
  {A.}~\bibnamefont {Ridolfi}},\ }\bibfield  {title} {\bibinfo {title} {{Pulsar
  J0453+1559: A Double Neutron Star System with a Large Mass Asymmetry}},\
  }\href {https://doi.org/10.1088/0004-637X/812/2/143} {\bibfield  {journal}
  {\bibinfo  {journal} {Astrophys. J.}\ }\textbf {\bibinfo {volume} {812}},\
  \bibinfo {pages} {143} (\bibinfo {year} {2015})},\ \Eprint
  {https://arxiv.org/abs/1509.08805} {arXiv:1509.08805 [astro-ph.HE]}
  \BibitemShut {NoStop}%
\bibitem [{\citenamefont {Suwa}\ \emph {et~al.}(2018)\citenamefont {Suwa},
  \citenamefont {Yoshida}, \citenamefont {Shibata}, \citenamefont {Umeda},\
  and\ \citenamefont {Takahashi}}]{Suwa:2018uni}%
  \BibitemOpen
  \bibfield  {author} {\bibinfo {author} {\bibfnamefont {Y.}~\bibnamefont
  {Suwa}}, \bibinfo {author} {\bibfnamefont {T.}~\bibnamefont {Yoshida}},
  \bibinfo {author} {\bibfnamefont {M.}~\bibnamefont {Shibata}}, \bibinfo
  {author} {\bibfnamefont {H.}~\bibnamefont {Umeda}},\ and\ \bibinfo {author}
  {\bibfnamefont {K.}~\bibnamefont {Takahashi}},\ }\bibfield  {title} {\bibinfo
  {title} {{On the minimum mass of neutron stars}},\ }\href
  {https://doi.org/10.1093/mnras/sty2460} {\bibfield  {journal} {\bibinfo
  {journal} {Mon. Not. Roy. Astron. Soc.}\ }\textbf {\bibinfo {volume} {481}},\
  \bibinfo {pages} {3305} (\bibinfo {year} {2018})},\ \Eprint
  {https://arxiv.org/abs/1808.02328} {arXiv:1808.02328 [astro-ph.HE]}
  \BibitemShut {NoStop}%
\bibitem [{\citenamefont {Stovall}\ \emph {et~al.}(2018)\citenamefont {Stovall}
  \emph {et~al.}}]{Stovall:2018ouw}%
  \BibitemOpen
  \bibfield  {author} {\bibinfo {author} {\bibfnamefont {K.}~\bibnamefont
  {Stovall}} \emph {et~al.},\ }\bibfield  {title} {\bibinfo {title} {{PALFA
  Discovery of a Highly Relativistic Double Neutron Star Binary}},\ }\href
  {https://doi.org/10.3847/2041-8213/aaad06} {\bibfield  {journal} {\bibinfo
  {journal} {Astrophys. J.}\ }\textbf {\bibinfo {volume} {854}},\ \bibinfo
  {pages} {L22} (\bibinfo {year} {2018})},\ \Eprint
  {https://arxiv.org/abs/1802.01707} {arXiv:1802.01707 [astro-ph.HE]}
  \BibitemShut {NoStop}%
\bibitem [{\citenamefont {Burgay}\ \emph {et~al.}(2003)\citenamefont {Burgay},
  \citenamefont {D'Amico}, \citenamefont {Possenti}, \citenamefont
  {Manchester}, \citenamefont {Lyne} \emph {et~al.}}]{Burgay:2003jj}%
  \BibitemOpen
  \bibfield  {author} {\bibinfo {author} {\bibfnamefont {M.}~\bibnamefont
  {Burgay}}, \bibinfo {author} {\bibfnamefont {N.}~\bibnamefont {D'Amico}},
  \bibinfo {author} {\bibfnamefont {A.}~\bibnamefont {Possenti}}, \bibinfo
  {author} {\bibfnamefont {R.}~\bibnamefont {Manchester}}, \bibinfo {author}
  {\bibfnamefont {A.}~\bibnamefont {Lyne}}, \emph {et~al.},\ }\bibfield
  {title} {\bibinfo {title} {{An Increased estimate of the merger rate of
  double neutron stars from observations of a highly relativistic system}},\
  }\href {https://doi.org/10.1038/nature02124} {\bibfield  {journal} {\bibinfo
  {journal} {Nature}\ }\textbf {\bibinfo {volume} {426}},\ \bibinfo {pages}
  {531} (\bibinfo {year} {2003})},\ \Eprint
  {https://arxiv.org/abs/astro-ph/0312071} {arXiv:astro-ph/0312071 [astro-ph]}
  \BibitemShut {NoStop}%
\bibitem [{\citenamefont {Abbott}\ \emph
  {et~al.}(2021{\natexlab{c}})\citenamefont {Abbott} \emph
  {et~al.}}]{LIGOScientific:2021psn}%
  \BibitemOpen
  \bibfield  {author} {\bibinfo {author} {\bibfnamefont {R.}~\bibnamefont
  {Abbott}} \emph {et~al.} (\bibinfo {collaboration} {LIGO Scientific, VIRGO,
  KAGRA}),\ }\bibfield  {title} {\bibinfo {title} {{The population of merging
  compact binaries inferred using gravitational waves through GWTC-3}},\ }\href
  {https://doi.org/10.48550/arXiv.2111.03634} {\bibfield  {journal} {\bibinfo
  {journal} {arXiv e-prints}\ ,\ \bibinfo {eid} {arXiv:2111.03634}} (\bibinfo
  {year} {2021}{\natexlab{c}})},\ \Eprint {https://arxiv.org/abs/2111.03634}
  {arXiv:2111.03634 [astro-ph.HE]} \BibitemShut {NoStop}%
\bibitem [{\citenamefont {Mandel}\ and\ \citenamefont
  {Broekgaarden}(2022)}]{Mandel:2021smh}%
  \BibitemOpen
  \bibfield  {author} {\bibinfo {author} {\bibfnamefont {I.}~\bibnamefont
  {Mandel}}\ and\ \bibinfo {author} {\bibfnamefont {F.~S.}\ \bibnamefont
  {Broekgaarden}},\ }\bibfield  {title} {\bibinfo {title} {{Rates of compact
  object coalescences}},\ }\href {https://doi.org/10.1007/s41114-021-00034-3}
  {\bibfield  {journal} {\bibinfo  {journal} {Living Rev. Rel.}\ }\textbf
  {\bibinfo {volume} {25}},\ \bibinfo {pages} {1} (\bibinfo {year} {2022})},\
  \Eprint {https://arxiv.org/abs/2107.14239} {arXiv:2107.14239 [astro-ph.HE]}
  \BibitemShut {NoStop}%
\bibitem [{\citenamefont {Smith}\ \emph {et~al.}(2016)\citenamefont {Smith},
  \citenamefont {Field}, \citenamefont {Blackburn}, \citenamefont {Haster},
  \citenamefont {Pürrer}, \citenamefont {Raymond},\ and\ \citenamefont
  {Schmidt}}]{Smith:2016qas}%
  \BibitemOpen
  \bibfield  {author} {\bibinfo {author} {\bibfnamefont {R.}~\bibnamefont
  {Smith}}, \bibinfo {author} {\bibfnamefont {S.~E.}\ \bibnamefont {Field}},
  \bibinfo {author} {\bibfnamefont {K.}~\bibnamefont {Blackburn}}, \bibinfo
  {author} {\bibfnamefont {C.-J.}\ \bibnamefont {Haster}}, \bibinfo {author}
  {\bibfnamefont {M.}~\bibnamefont {Pürrer}}, \bibinfo {author} {\bibfnamefont
  {V.}~\bibnamefont {Raymond}},\ and\ \bibinfo {author} {\bibfnamefont
  {P.}~\bibnamefont {Schmidt}},\ }\bibfield  {title} {\bibinfo {title} {{Fast
  and accurate inference on gravitational waves from precessing compact
  binaries}},\ }\href {https://doi.org/10.1103/PhysRevD.94.044031} {\bibfield
  {journal} {\bibinfo  {journal} {Phys. Rev.}\ }\textbf {\bibinfo {volume}
  {D94}},\ \bibinfo {pages} {044031} (\bibinfo {year} {2016})},\ \Eprint
  {https://arxiv.org/abs/1604.08253} {arXiv:1604.08253 [gr-qc]} \BibitemShut
  {NoStop}%
\bibitem [{\citenamefont {Smith}\ \emph {et~al.}(2020)\citenamefont {Smith},
  \citenamefont {Ashton}, \citenamefont {Vajpeyi},\ and\ \citenamefont
  {Talbot}}]{Smith:2020}%
  \BibitemOpen
  \bibfield  {author} {\bibinfo {author} {\bibfnamefont {R.~J.~E.}\
  \bibnamefont {Smith}}, \bibinfo {author} {\bibfnamefont {G.}~\bibnamefont
  {Ashton}}, \bibinfo {author} {\bibfnamefont {A.}~\bibnamefont {Vajpeyi}},\
  and\ \bibinfo {author} {\bibfnamefont {C.}~\bibnamefont {Talbot}},\
  }\bibfield  {title} {\bibinfo {title} {{Massively parallel Bayesian inference
  for transient gravitational-wave astronomy}},\ }\href
  {https://doi.org/10.1093/mnras/staa2483} {\bibfield  {journal} {\bibinfo
  {journal} {Monthly Notices of the Royal Astronomical Society}\ }\textbf
  {\bibinfo {volume} {498}},\ \bibinfo {pages} {4492} (\bibinfo {year}
  {2020})}\BibitemShut {NoStop}%
\bibitem [{\citenamefont {Ashton}\ \emph {et~al.}(2019)\citenamefont {Ashton}
  \emph {et~al.}}]{Ashton:2018jfp}%
  \BibitemOpen
  \bibfield  {author} {\bibinfo {author} {\bibfnamefont {G.}~\bibnamefont
  {Ashton}} \emph {et~al.},\ }\bibfield  {title} {\bibinfo {title} {{BILBY: A
  user-friendly Bayesian inference library for gravitational-wave astronomy}},\
  }\href {https://doi.org/10.3847/1538-4365/ab06fc} {\bibfield  {journal}
  {\bibinfo  {journal} {Astrophys. J. Suppl.}\ }\textbf {\bibinfo {volume}
  {241}},\ \bibinfo {pages} {27} (\bibinfo {year} {2019})},\ \Eprint
  {https://arxiv.org/abs/1811.02042} {arXiv:1811.02042 [astro-ph.IM]}
  \BibitemShut {NoStop}%
\bibitem [{\citenamefont {Abbott}\ \emph
  {et~al.}(2021{\natexlab{d}})\citenamefont {Abbott} \emph
  {et~al.}}]{LIGOScientific:2019lzm}%
  \BibitemOpen
  \bibfield  {author} {\bibinfo {author} {\bibfnamefont {R.}~\bibnamefont
  {Abbott}} \emph {et~al.} (\bibinfo {collaboration} {LIGO Scientific,
  Virgo}),\ }\bibfield  {title} {\bibinfo {title} {{Open data from the first
  and second observing runs of Advanced LIGO and Advanced Virgo}},\ }\href
  {https://doi.org/10.1016/j.softx.2021.100658} {\bibfield  {journal} {\bibinfo
   {journal} {SoftwareX}\ }\textbf {\bibinfo {volume} {13}},\ \bibinfo {pages}
  {100658} (\bibinfo {year} {2021}{\natexlab{d}})},\ \Eprint
  {https://arxiv.org/abs/1912.11716} {arXiv:1912.11716 [gr-qc]} \BibitemShut
  {NoStop}%
\bibitem [{\citenamefont {Romani}\ \emph {et~al.}(2022)\citenamefont {Romani},
  \citenamefont {Kandel}, \citenamefont {Filippenko}, \citenamefont {Brink},\
  and\ \citenamefont {Zheng}}]{Romani:2022}%
  \BibitemOpen
  \bibfield  {author} {\bibinfo {author} {\bibfnamefont {R.~W.}\ \bibnamefont
  {Romani}}, \bibinfo {author} {\bibfnamefont {D.}~\bibnamefont {Kandel}},
  \bibinfo {author} {\bibfnamefont {A.~V.}\ \bibnamefont {Filippenko}},
  \bibinfo {author} {\bibfnamefont {T.~G.}\ \bibnamefont {Brink}},\ and\
  \bibinfo {author} {\bibfnamefont {W.}~\bibnamefont {Zheng}},\ }\bibfield
  {title} {\bibinfo {title} {Psr j0952-0607: The fastest and heaviest known
  galactic neutron star},\ }\href {https://doi.org/10.3847/2041-8213/ac8007}
  {\bibfield  {journal} {\bibinfo  {journal} {The Astrophysical Journal
  Letters}\ }\textbf {\bibinfo {volume} {934}},\ \bibinfo {pages} {L17}
  (\bibinfo {year} {2022})}\BibitemShut {NoStop}%
\bibitem [{\citenamefont {Narikawa}\ and\ \citenamefont
  {Uchikata}(2022)}]{Narikawa:2022saj}%
  \BibitemOpen
  \bibfield  {author} {\bibinfo {author} {\bibfnamefont {T.}~\bibnamefont
  {Narikawa}}\ and\ \bibinfo {author} {\bibfnamefont {N.}~\bibnamefont
  {Uchikata}},\ }\bibfield  {title} {\bibinfo {title} {{Follow-up analyses of
  the binary-neutron-star signals GW170817 and GW190425 by using post-Newtonian
  waveform models}},\ }\href {https://doi.org/10.1103/PhysRevD.106.103006}
  {\bibfield  {journal} {\bibinfo  {journal} {Phys. Rev. D}\ }\textbf {\bibinfo
  {volume} {106}},\ \bibinfo {pages} {103006} (\bibinfo {year} {2022})},\
  \Eprint {https://arxiv.org/abs/2205.06023} {arXiv:2205.06023 [gr-qc]}
  \BibitemShut {NoStop}%
\bibitem [{\citenamefont {De}\ \emph {et~al.}(2018)\citenamefont {De},
  \citenamefont {Finstad}, \citenamefont {Lattimer}, \citenamefont {Brown},
  \citenamefont {Berger},\ and\ \citenamefont {Biwer}}]{De:2018uhw}%
  \BibitemOpen
  \bibfield  {author} {\bibinfo {author} {\bibfnamefont {S.}~\bibnamefont
  {De}}, \bibinfo {author} {\bibfnamefont {D.}~\bibnamefont {Finstad}},
  \bibinfo {author} {\bibfnamefont {J.~M.}\ \bibnamefont {Lattimer}}, \bibinfo
  {author} {\bibfnamefont {D.~A.}\ \bibnamefont {Brown}}, \bibinfo {author}
  {\bibfnamefont {E.}~\bibnamefont {Berger}},\ and\ \bibinfo {author}
  {\bibfnamefont {C.~M.}\ \bibnamefont {Biwer}},\ }\bibfield  {title} {\bibinfo
  {title} {{Tidal Deformabilities and Radii of Neutron Stars from the
  Observation of GW170817}},\ }\href
  {https://doi.org/10.1103/PhysRevLett.121.091102} {\bibfield  {journal}
  {\bibinfo  {journal} {Phys. Rev. Lett.}\ }\textbf {\bibinfo {volume} {121}},\
  \bibinfo {pages} {091102} (\bibinfo {year} {2018})},\ \Eprint
  {https://arxiv.org/abs/1804.08583} {arXiv:1804.08583 [astro-ph.HE]}
  \BibitemShut {NoStop}%
\bibitem [{\citenamefont {Capano}\ \emph {et~al.}(2020)\citenamefont {Capano},
  \citenamefont {Tews}, \citenamefont {Brown}, \citenamefont {Margalit},
  \citenamefont {De}, \citenamefont {Kumar}, \citenamefont {Brown},
  \citenamefont {Krishnan},\ and\ \citenamefont {Reddy}}]{Capano:2019eae}%
  \BibitemOpen
  \bibfield  {author} {\bibinfo {author} {\bibfnamefont {C.~D.}\ \bibnamefont
  {Capano}}, \bibinfo {author} {\bibfnamefont {I.}~\bibnamefont {Tews}},
  \bibinfo {author} {\bibfnamefont {S.~M.}\ \bibnamefont {Brown}}, \bibinfo
  {author} {\bibfnamefont {B.}~\bibnamefont {Margalit}}, \bibinfo {author}
  {\bibfnamefont {S.}~\bibnamefont {De}}, \bibinfo {author} {\bibfnamefont
  {S.}~\bibnamefont {Kumar}}, \bibinfo {author} {\bibfnamefont {D.~A.}\
  \bibnamefont {Brown}}, \bibinfo {author} {\bibfnamefont {B.}~\bibnamefont
  {Krishnan}},\ and\ \bibinfo {author} {\bibfnamefont {S.}~\bibnamefont
  {Reddy}},\ }\bibfield  {title} {\bibinfo {title} {{Stringent constraints on
  neutron-star radii from multimessenger observations and nuclear theory}},\
  }\href {https://doi.org/10.1038/s41550-020-1014-6} {\bibfield  {journal}
  {\bibinfo  {journal} {Nature Astron.}\ }\textbf {\bibinfo {volume} {4}},\
  \bibinfo {pages} {625} (\bibinfo {year} {2020})},\ \Eprint
  {https://arxiv.org/abs/1908.10352} {arXiv:1908.10352 [astro-ph.HE]}
  \BibitemShut {NoStop}%
\bibitem [{\citenamefont {Kunert}\ \emph {et~al.}(2022)\citenamefont {Kunert},
  \citenamefont {Pang}, \citenamefont {Tews}, \citenamefont {Coughlin},\ and\
  \citenamefont {Dietrich}}]{Kunert:2021hgm}%
  \BibitemOpen
  \bibfield  {author} {\bibinfo {author} {\bibfnamefont {N.}~\bibnamefont
  {Kunert}}, \bibinfo {author} {\bibfnamefont {P.~T.~H.}\ \bibnamefont {Pang}},
  \bibinfo {author} {\bibfnamefont {I.}~\bibnamefont {Tews}}, \bibinfo {author}
  {\bibfnamefont {M.~W.}\ \bibnamefont {Coughlin}},\ and\ \bibinfo {author}
  {\bibfnamefont {T.}~\bibnamefont {Dietrich}},\ }\bibfield  {title} {\bibinfo
  {title} {{Quantifying modeling uncertainties when combining multiple
  gravitational-wave detections from binary neutron star sources}},\ }\href
  {https://doi.org/10.1103/PhysRevD.105.L061301} {\bibfield  {journal}
  {\bibinfo  {journal} {Phys. Rev. D}\ }\textbf {\bibinfo {volume} {105}},\
  \bibinfo {pages} {L061301} (\bibinfo {year} {2022})},\ \Eprint
  {https://arxiv.org/abs/2110.11835} {arXiv:2110.11835 [astro-ph.HE]}
  \BibitemShut {NoStop}%
\end{thebibliography}%
\newpage
\section*{Supplemental Material}
\subsection{Injection Parameters}

We study injected signals generated with the \textsc{IMRPhenomD\_NRTidalv2} approximant~\cite{Dietrich:2019kaq}.
The injected component masses $M_{1,2}$ are drawn from a Gaussian distribution $\mathcal{N}(\mu=1.33, \sigma=0.09)$ that is characteristic for galactic BNS systems~\cite{Ozel:2016oaf}.
They are uniformly distributed in a comoving volume with a distance cutoff at \SI{200}{Mpc}.
In a larger random sample of 1000 systems within \SI{500}{Mpc}, the average chirp mass of observable binaries settles at this distance near the injected distribution's mean, indicating that the volume is sufficiently large to characterize the underlying distribution.
We further limit our analysis to systems with a signal-to-noise ratio (SNR) above 30. This value is sufficiently high to expect measurements of significant tidal contributions without introducing a bias towards higher masses, where tidal effects would again become less prominent.
For a given mass $M$, the respective EOS fully determines radius $R$ and tidal deformability $\Lambda:=\frac{2k_2 R^5}{3M^5}$, where 
$k_2$ denotes the tidal Love number~\cite{Hinderer:2007mb}. 
Since only \textLambda\ is prominent in the waveform, we base our EOS selection on its distribution at a relatively low \SI{1.2}{M_\odot}.
This mass yet is firmly supported by observations and neutron star formation theories~\cite{Martinez:2015mya,Suwa:2018uni}.
\Cref{fig:m-r-lam-prior} shows how the parameter spaces of the EOS sets largely overlap at high masses, corresponding to core densities far beyond the breakdown of the chiral EFT approach. 
We then choose to inject an EOS from each distribution's 50th percentile which has a TOV mass closest to a fiducial value of \SI{2.2}{M_\odot}.
The dimensionless aligned spins $\chi_{1,2}$ are constrained to a uniform distribution subject to $|\chi_i|<0.05$, as implied for realistic sources of NS mergers~\cite{Stovall:2018ouw,Burgay:2003jj}.
We ultimately leave the sky location, inclination angle $\theta_{JN}$, orbital phase at coalescence $\phi$, and polarisation angle $\psi$ totally unconstrained.
Subject to the population model under consideration, the detection of 20 such systems will amount to at least two years of observation at ET~\cite{LIGOScientific:2021psn,Mandel:2021smh}.

\subsection{Bayesian Inference}
We study the resulting effects in the framework of Bayesian inference, using the EOS as a sampling parameter that is constrained by tidal terms in the observed waveform.
This makes use of Bayes' theorem \begin{align}
    \label{eq:Bayes}
    p (\theta|d)= \frac{L(d|\theta) \pi(\theta)}{\mathcal{Z}(d)}, \text{ with }\\
    \mathcal{Z}(d)=\int_\Theta L(d|\theta) \pi(\theta) \, {\rm d}\theta,
\end{align}
to determine a posterior distribution $p (\theta|d)$ of the multi-dimensional parameter space $\Theta$ that characterizes an event's GW strain.
We reweight a parameter set's prior probability $ \pi(\theta)$ by the likelihood $L$ that it is the cause of the observed data $d$.

\begin{table}[thb]
    \centering
    \begin{tabular}{clrlrl}
    &   parameter  &   \multicolumn{2}{c}{symbol} & \multicolumn{2}{c}{prior bounds} \\
      \hline
       \multirow{6}{8.5pt}{\rotatebox{90}{observational}}& luminosity distance [Mpc] &&$d_{L}$ & 5 &-- 500 \\
       & inclination &$\cos$& $\theta_{JN}$& -1 &-- 1\\
      & phase [rad] &&$\phi$ & 0 &-- $2\pi$\\
       & polarisation [rad] &&$\psi$ & 0 &-- $\pi$ \\
        & right ascension [rad] && \textalpha & 0 & -- $2\pi$\\
        & declination [rad] && \textdelta & $-\pi$ & -- $\pi$\vspace{0.1cm}\\
      \multirow{5}{8.5pt}{\rotatebox{90}{orbital}} & chirp mass $[\rm M_\odot] $ &&$\mathcal{M}$ &  \textit{1.20} &-- \textit{1.30}* \\
      & source chirp mass $[\rm M_\odot] $ &&$\mathcal{M}_s$ &  1.15&-- 1.30* \\
       & mass ratio && $q$ & 0.125 &-- 1\\
       &source comp. mass $[\rm M_\odot]$ && $M_{i,s}$ & \textit{ \textgreater 0.5}\\
        & aligned component spin & &$\chi_i$ & -0.15&-- 0.15 \vspace{0.1cm}\\
        hyper & Equation of State &&EOS & 1 &-- 3000
    \end{tabular}
    \caption[GW Sampling Parameters in the ET-Analysis]
    {\textbf{GW Sampling Parameters in the ET-Analysis}:
    Most priors are uniform within given bounds. 
    The declination \textdelta\ is uniform in cosine, and the luminosity distance $d_{\rm L}$ is uniform within a co-moving volume of the specified dimension. 
    The EOS prior is weighted by the ability to support massive pulsars. 
    Prior ranges in italics indicate constraints that are not used as sampling parameters. 
    Starred priors are adjusted to the injected signal.}
    \label{tab:ET_priors}
\end{table}
The sampling parameters of our injection study are given in Table \ref{tab:ET_priors}. 
These include observational parameters (e.g., luminosity distance $d_L$, phase, inclination angles of the merger) and intrinsic binary parameters (e.g., chirp mass~$\mathcal{M}$, mass ratio, tilts).
We considerably extend the range above the injection distribution for the luminosity distance and aligned spins in order to avoid boundary effects from the prior distribution.
We note, though, that this comes at the cost of a bias towards unequal mass ratios in the parameter estimation.
We weight the EOSs conservatively by their ability to support the most firmly established pulsar masses~\cite{Antoniadis:2013pzd,Fonseca:2021wxt,Arzoumanian:2017puf}.
In order to reduce the significant computational cost, we apply a Reduced-Order-Quadrature (ROQ) rule~\cite{Smith:2016qas}. 
This requires limiting the chirp mass space to \SI{0.1}{M_\odot} intervals.
As we sample over chirp masses in the source frame, we invoke a corresponding prior adapted to the injected signal in order to avoid computational issues. 
Since the chirp mass is by far the most accurately measured quantity, the prior is still wide enough to avoid the introduction of prior-driven artefacts in the parameter estimation.

We use \textsc{parallel-bilby}~\cite{Smith:2020}, an efficient parallelisation package relying on nested sampling routines from \textsc{bilby}~\cite{Skilling:2006gxv,Ashton:2018jfp}, to obtain the evidence for either EOS set.
Nested sampling algorithms aim at calculating $\mathcal{Z}$ and yield the posterior \textit{en passant}.
Employing some minor modifications to \textsc{parallel-bilby} and \textsc{bilby}, we can sample over EOSs from the respective set.
Radii and tidal deformabilities then follow uniquely from the component masses.
The likelihood evaluations follow the usual matched filter approach employing the \textsc{IMRPhenomD\_NRTidalv2} approximant to efficiently generate waveforms including tidal effects.
To further reduce computational costs, we perform inference in the frequency range \SI{30}{Hz} to \SI{2048}{Hz}.
We use 2048 live points for nested sampling.
The selection of the \textsc{IMRPhenomD\_NRTidalv2} approximant is driven by efficiency and robustness considerations~\cite{Dietrich:2019kaq}.
In this setting, each inference run requires about 80,000 hours of computing time.

We can use the evidence $\mathcal{Z}$ for model selection because a higher evidence can only be achieved by the more complex one (i.e. with a less compact prior $\pi$) among two competing models if it matches the data significantly better.
Treating both models of 3N interaction as a priori equally likely, preference is expressed by the Bayes factor $\bf= \mathcal{Z}_{\ve} /\mathcal{Z}_{\rm TPE}$.
The evidence for either model explaining a suite of $N$ independent observations is given by
\begin{align}
    \mathcal{Z}&= \int \prod_{i=1}^N {L}_i(\theta_i, {\rm EOS}_i) \pi(\theta_i, {\rm EOS}_i) \,\dift\theta_i \dift{\rm EOS}_i  \\
    &= \prod_{i=1}^N \int  {L}_i(\theta_i, {\rm EOS}_i) \pi(\theta_i, {\rm EOS}_i) \,\dift\theta_i \dift{\rm EOS}_i  \\
    &= \prod_{i=1}^N \mathcal{Z}_i.\label{eq:usual_bayes}
\end{align}
We have subsumed all system parameters besides the EOS in $\theta_i$.
The corresponding Bayes factor \begin{align}
    \bf= \prod_{i=1}^N \frac{\mathcal{Z}_{\ve,i}}{\mathcal{Z}_{{\rm TPE},i}}
\end{align} then expresses the statistical support for the notion that the $N$ systems are characterised by the \mve\ description instead of TPE.

\begin{figure}[t]
    \centering
    \includegraphics[width=0.48\textwidth]{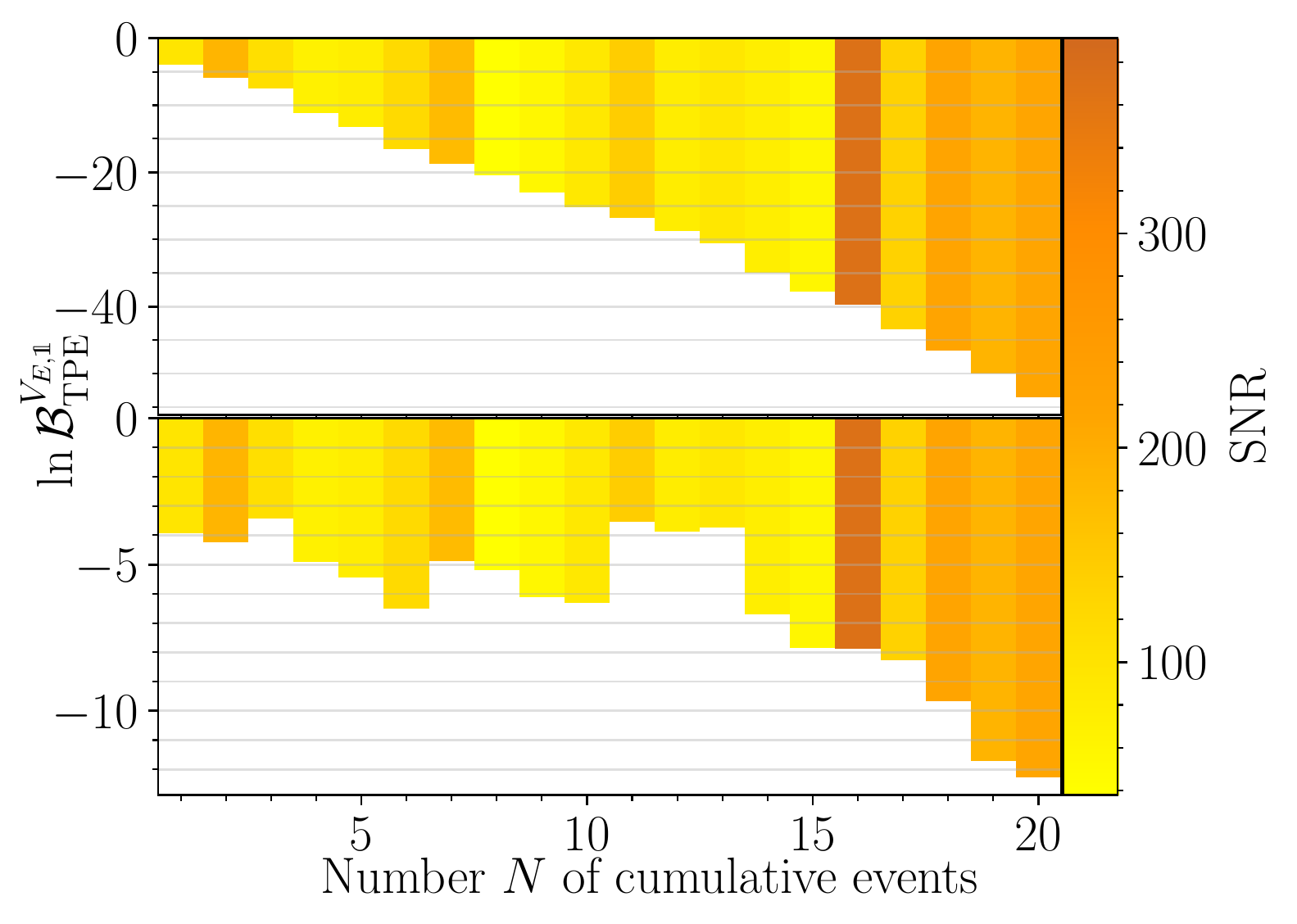}
    \includegraphics[width=0.48\textwidth]{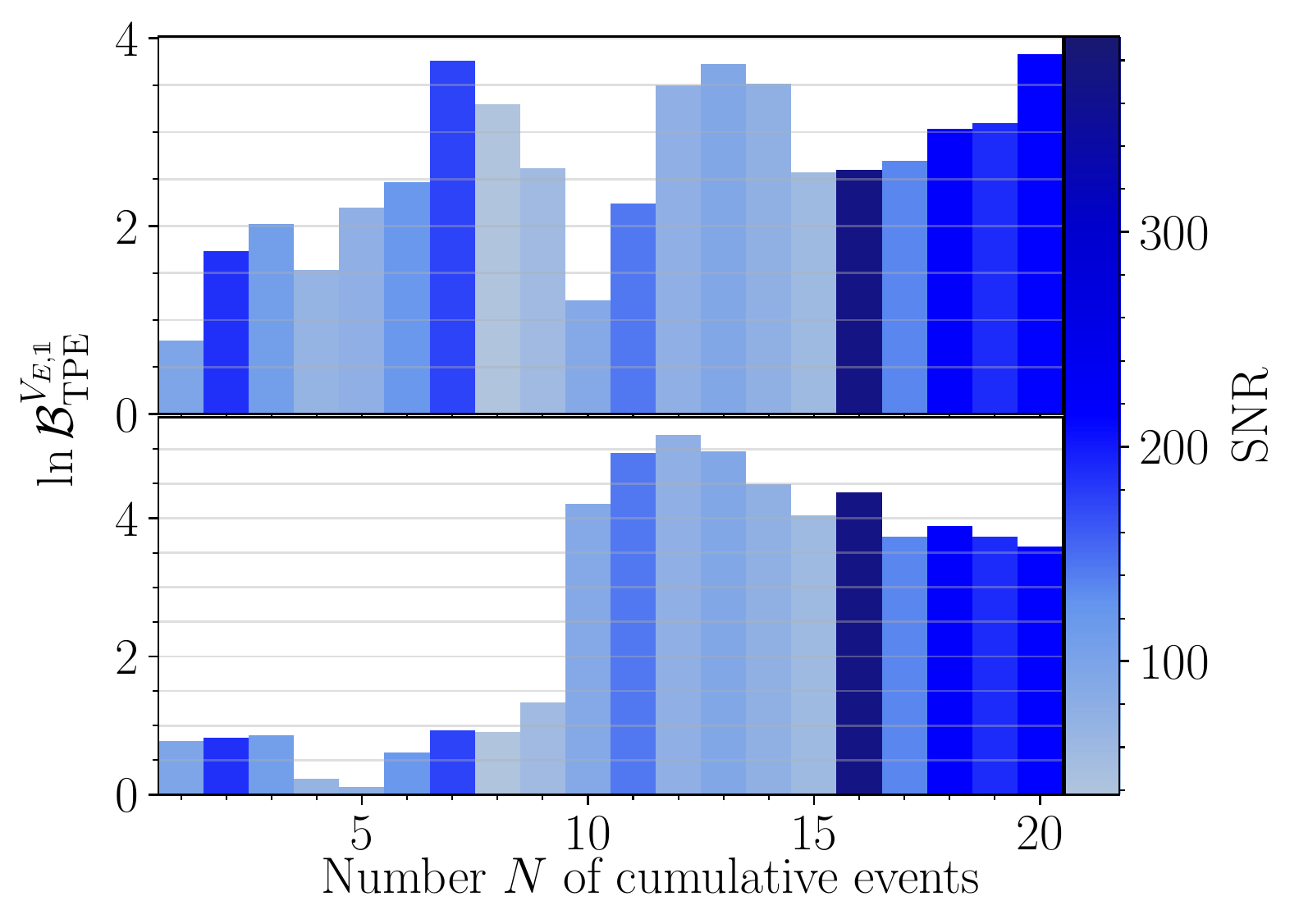}
    \caption{\textbf{Comparison of Bayes Factors:} 
    Bayes factors as in Fig.~\ref{fig:main_result}, using eq.~\ref{eq:usual_bayes} in the upper and eq.~\ref{eq:alt_bayes} in the lower part.
    Note the different scales.
    For the TPE (top) injection, the model preference becomes less decisive and the inclusion of some runs favors the \mve\ model when assuming that all observations result from the same EOS.
    For the \mve\ injection (bottom), the overall model preference remains nearly constant, while the contribution of some runs varies greatly.
    }
    \label{fig:alt_result}
\end{figure}

In principle, one could also include the fact that all observed systems should be explained by exactly one EOS.
This leads to the alternative expression for the evidence
\begin{align}
    \mathcal{Z}=& \int \prod_{i=1}^N {L}_i(\theta_i, {\rm EOS}_i) \pi(\theta_i, {\rm EOS}_i) \nonumber\\
    &\cdot\delta({\rm EOS}_i - {\rm EOS}_1) \,\dift\theta_i \dift{\rm EOS}_i  \\
    =& \int \prod_{i=1}^N {L}_i(\theta_i, {\rm EOS}) \pi(\theta_i, {\rm EOS}) \,\dift\theta_i \dift{\rm EOS}  \\
    =& \int \prod_{i=1}^N \mathcal{Z}_i p_i({\rm EOS}) \,\dift{\rm EOS}  \\
    =& \prod_{j=1}^N \mathcal{Z}_j\int \prod_{i=1}^N p_i({\rm EOS}) \,\dift{\rm EOS},\label{eq:alt_bayes}
\end{align}
with $p_i({\rm EOS})$ denoting the posterior of inference run $i$ marginalised over all parameters but the EOS.
However, this prescription does not reflect the construction of our EOS sets that are meant to convey current modelling uncertainties in chiral EFT. 
The EOS parameter space therefore leads to unequal prior densities in \textLambda.
Consider, for instance, a segment of \textLambda\ space at relatively low mass that is only approximately met by a single TPE EOS, whereas multiple \mve\ EOS provide similarly good agreement with observations.
This would naturally happen if \mve\ describes the true EOS, independent of the total number of EOSs in each set.
After some mergers with near solar mass NSs, eq.~\ref{eq:alt_bayes} would still suggest model preference for the inappropriate TPE description because it matches the expectation of a single true EOS better.
We see this effect in Fig.~\ref{fig:alt_result}, massively reducing model preference in case of the TPE injection.
Nevertheless, we use the assumption that a single EOS should be responsible for all observations in the related estimate on $R_{1.4}$ and $\Lambda_{1.4}$.
As these necessarily imply an approximation based on the EOS, we employ a joint EOS posterior.
Its distribution $p({\rm EOS})$ after observing $N$ systems is given by
\begin{align*}
    p({\rm EOS})= const.\cdot \frac{\prod_{i=1}^N     p_i({\rm EOS})}{\pi({\rm EOS})^{N-1}},
\end{align*}
with $p_i({\rm EOS})$ denoting the posterior distribution obtained from the $i$-th event.

\subsection{Re-analysis of GW170817}
\begin{table}[tbhp]
    \centering
    \begin{tabular}{clrlrl}
    &   parameter  &   \multicolumn{2}{c}{symbol} & \multicolumn{2}{c}{prior bounds} \\
      \hline
       \multirow{6}{8.5pt}{\rotatebox{90}{observational}}& lum. distance [Mpc] &&$d_{\rm L}$ & 1 &-- 75 \\
       & inclination &$\cos$& $\theta_{JN}$& -1 &-- 1\\
      & phase [rad] &&$\phi$ & 0 &-- $2\pi$\\
       & polarisation [rad] &&$\psi$ & 0 &-- $\pi$ \\
        & right ascension [rad]&& \textalpha &  3.44616 & (exact)\\
        & declination [rad]& & \textdelta &  -0.408084 & (exact) \vspace{0.1cm}\\
      \multirow{4}{8.5pt}{\rotatebox{90}{orbital}} & chirp mass $[M_\odot] $ &&$\mathcal{M}$ &  1.18 &-- 1.21 \\
       & mass ratio && $q$ & 0.125 &-- 1\\
       &component mass $[M_\odot]$ && $M_{i}$ & \textit{ \textgreater 1.0}\\
        & aligned component spin & &$\chi_i$ & -0.15&-- 0.15 \vspace{0.1cm}\\
        hyper & Equation of State &&EOS & 1 &-- 3000
    \end{tabular}
    \caption[GW Sampling Parameters in GW170817-Analysis]
    {\textbf{GW Sampling Parameters in GW170817-Analysis}: 
    Most priors are uniform within given bounds. 
    Luminosity distance $d_{\rm L}$ is uniform within a comoving volume of the specified radial dimension. 
    The EOS prior is weighted by the ability to support mass constraints from high-mass pulsars, NICER observations, and the kilonova observations that suggested the formation of a hypermassive NS. 
    The component mass prior indicates a constraint that is not used as a sampling parameter. }
    \label{tab:gw_priors}
\end{table}
For a reanalysis of GW170817~\cite{LIGOScientific:2019lzm}, we use the available information on the GRB afterglow and kilonova, motivating the modified prior distribution given in Table \ref{tab:gw_priors}. 
We also use a more informative EOS prior that is weighted by minimum mass constraints from precise pulsar observations~\cite{Romani:2022,Arzoumanian:2017puf,Antoniadis:2013pzd,Fonseca:2021wxt}, evidence for the formation of a hypermassive neutron star in the merger~\cite{Rezzolla:2017aly,Margalit:2017dij}, and NICER analysis of millisecond pulsars~\cite{Riley:2019yda,Riley:2021pdl,Miller:2019cac,Miller:2021qha}.
Each measurement is assumed to be subject to Gaussian errors characterized by the respectively published uncertainty.

\begin{figure*}
    \centering
    \includegraphics[width=0.97\textwidth]{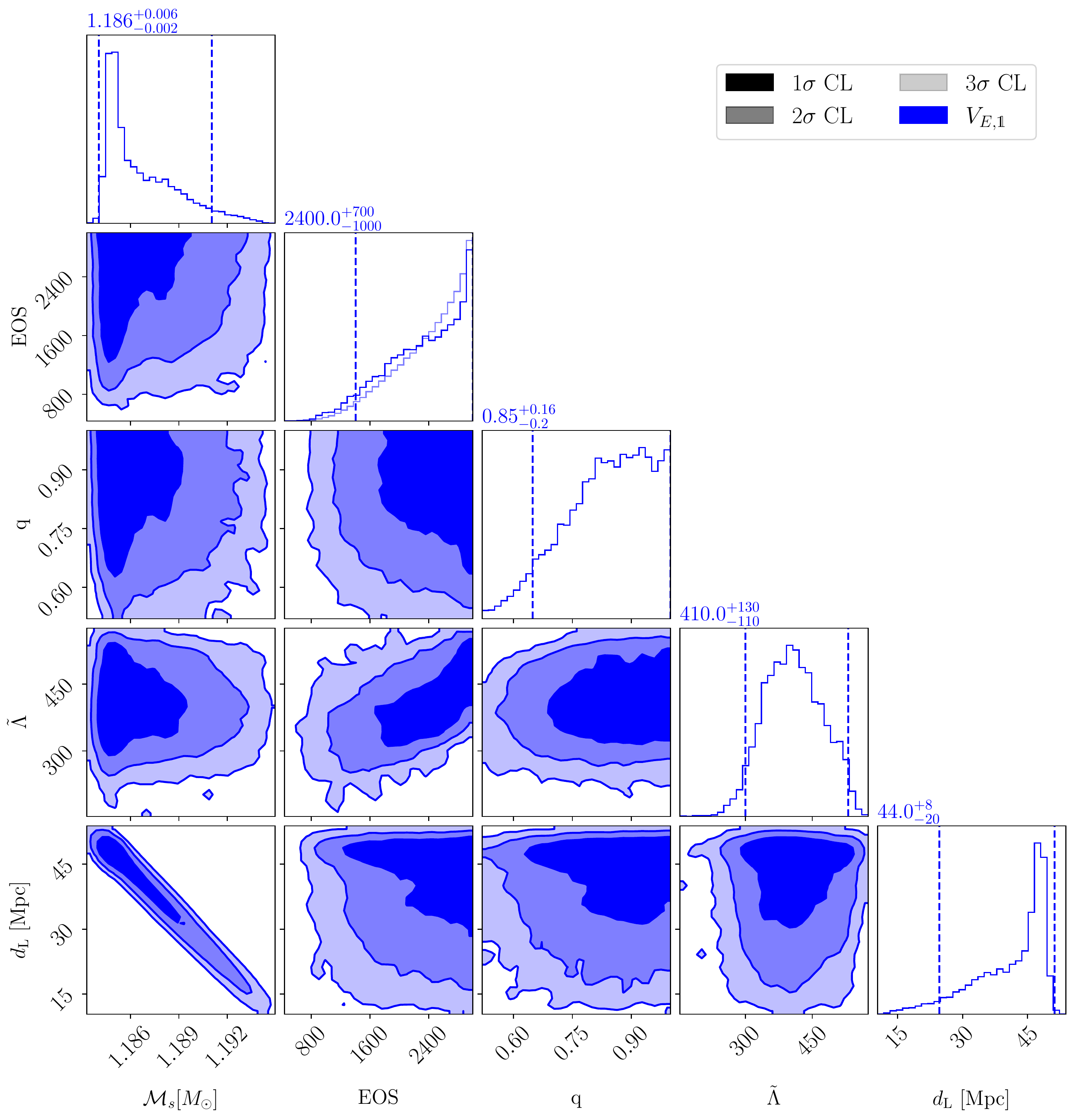}
    \caption[Important System Parameters for GW170817]
    {\textbf{Important System Parameters for GW170817}:
    We show a corner plot of (in reading direction) chirp mass, EOS index, mass ratio, tidal deformability, and luminosity distance. 
    The blue contours represent the \mve~recovery. 
    A TPE recovery is not included because the low information gain in the EOS posterior (as indicated by the marginal deviation from the prior, shown in the corresponding histogram as a faint line) suggests that no constraints can be won. Dashed lines in the top histograms mark the 90\% CI, contours indicate 1-, 2-, and 3-\textsigma\ confidence levels in the 2D-histograms.} 
    \label{fig:GW170817_corner}
\end{figure*}
Fig. \ref{fig:GW170817_corner} displays the recovered spread of several key parameters which are consistent with the original findings and recent reanalysis~\cite{LIGOScientific:2017qsa,Narikawa:2022saj}. 
The luminosity distance peaks sharply near \SI{46}{Mpc}, matching the spread in source chirp mass.
This is slightly lower than originally reported.
We can associate this effect with the wide spread of mass ratios, falling even below 0.6.
The low mass ratios correspond to the extended prior range for the aligned spin components in comparison with Ref. \cite{LIGOScientific:2017qsa} that considered the case $|{\chi_i}|<0.05$.
Since we find the spins (not shown) to deviate only slightly from the prior distribution and to be strongly anti-correlated, we conclude that there is no evidence for significant spin effects.
The tidal deformability is relatively tightly constrained, falling way below the limits in nuclear-physics agnostic analysis of the original discovery~\cite{Abbott:2018exr,De:2018uhw,Abbott:2018wiz} and matching findings of $80\leq\Tilde{\Lambda}\leq 580$ in a similar chiral EFT framework~\cite{Tews:2018iwm}.
This is a prior-driven conclusion, though, and we find that the EOS distribution has hardly relaxed from the prior.
This is due to the fact that GW170817 and chiral EFT up to $2\rho_{\rm sat}$ provide similar information on the EOS~\cite{Capano:2019eae}.
Since this run at the upper limit of plausible deformabilities is uninformative, no better constraints can be expected from a TPE recovery on these data.
This analysis does, therefore, suggest no preference for any particular realisation of 3N interactions.
Other GW detections with neutron stars have so far proven even less informative with respect to tidal effects.
Further measurements of NS mergers are expected in the next observing runs, but current population models make it unlikely that these include signals that are considerably stronger than GW170817.

\subsection{Posterior Validation}
\label{ssec:post_val}
Our analysis comes with some caveats that we address in the following.
Ref.~\cite{Kunert:2021hgm} has shown in a comparable framework how systematic errors resulting from approximations in available waveform approximants affect the determination of tidal effects.
Given the greatly increased sensitivity of ET and the prospect of advances in waveform modelling in the upcoming years, we are optimistic that these uncertainties will be reduced significantly when analyzing future detections.
We further based our analysis on observations above \SI{30}{Hz} where tidal effects begin to contribute.
Inference on the full detection band would have further increased the high computational cost of this study by orders of magnitude.
The mass and spin parameters, however, are best determined at \SI{5}{Hz} to \SI{9}{Hz}~\cite{Dietrich:2020eud}.
Measuring them with high precision in this range would naturally constrain the inference of tidal parameters, too.
Similarly, a signal recorded by a GW detector network or even identified in optical counterparts would constrain the sky localisation much tighter.
To mimic these effects, we re-analyse a signal with particularly poor parameter estimation in the \mve\ injection under the assumption, that a) mass parameters and sky localisation were tightly constrained -- as expected from a full bandwidth detection in a GW detector network --, and that b) the luminosity distance was precisely known, as expected from the identification of the host galaxy to an EM counterpart.

\begin{figure*}
    \centering
    \includegraphics[width=0.75\textwidth]{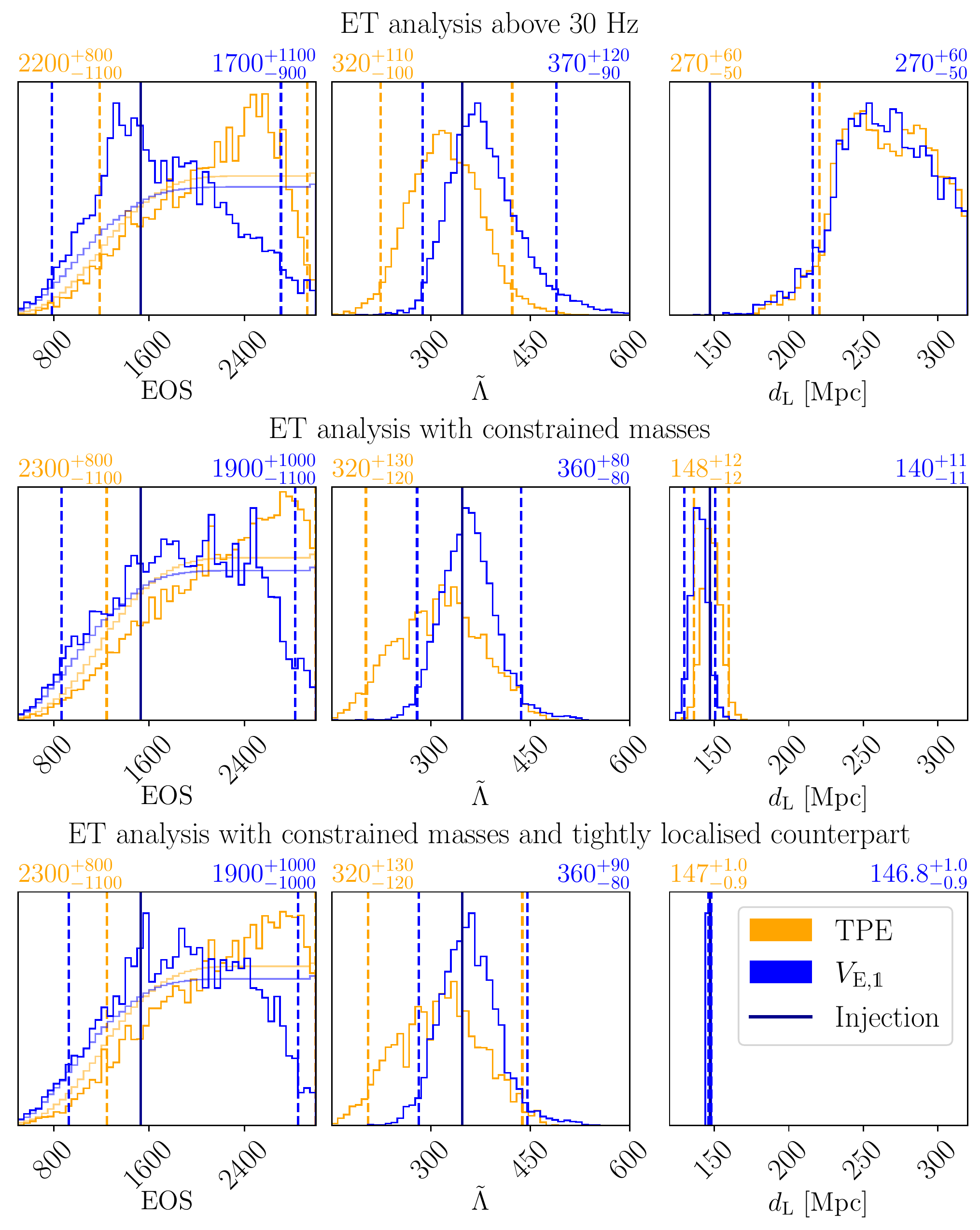}
    \caption[Selected posteriors with adapted priors]
    {\label{fig:prior_var}\textbf{Selected posteriors with adapted priors:} We show the posteriors for the EOS indices (left), the associated tidal deformability (center), and luminosity distance (right) in our original set-up (top), with tightly constrained mass parameters as expected from realistic network operation (middle), and with known distance as expected from the localisation of an EM counterpart (bottom). 
    The fainter lines in the EOS plots indicate the indices' prior weight.
    Note how the distance estimate and the tidal description improve when assuming knowledge on the mass parameters.
    Further limiting the luminosity distance does not improve the quality of other parameters.}
\end{figure*}
Fig. \ref{fig:prior_var} shows that the first option does indeed lead to a more plausible description of the tidal effects.
The major distance overestimate in our original analysis drives the EOS sets to EOSs with a lower tidal deformability to counter the underestimate in the (redshifted) source frame mass parameters. 
These are associated with more more compact neutron stars that typically have a lower TOV limit.
Because our prior penalizes low TOV limits, good waveform fits had previously worse prior support, particularly for \mve. 
Properly identifying the detected mass within narrow margins of \SI{0.01}{M_\odot} removes this source of uncertainty considerably and resolves the erroneous model preference for TPE against the injected \mve\ EOS.
Constraining the luminosity distance even further in the second step does not improve the estimation of other parameters.

Moreover, we reanalyze this signal with a different EOS prior that penalizes TOV limits above $2.16^{+0.17}_{-0.15}\,\rm M_\odot$~\cite{Rezzolla:2017aly}.
This step does not significantly improve the \mve\ parameter estimation, but leads to a modified EOS posterior.
The TPE sampling, in contrast, does not converge within acceptable runtime because the adjusted prior effectively outlaws EOSs that previously allowed suitable waveform descriptions.
This makes it much harder for the nested sampling algorithm to find parameters with better likelihood.
Enforcing convergence by allocating significantly more computing resources would certainly have removed the model preference for TPE against the injection.
This supports our conclusion that realistic GW detections in the ET era with improved priors from upcoming detections will be capable of quickly distinguishing 3N interactions.

\end{document}